\documentclass[11pt]{elsarticle}
\usepackage{amsmath,amsthm,amssymb}
\usepackage{latexsym}
\usepackage{color}
\usepackage{url}
\usepackage{epsfig}
\usepackage{array}
\usepackage{tabularx}
\usepackage{subfigure}
\journal{Computerized Medical Imaging and Graphics}

\usepackage{graphicx}
\usepackage{amsmath}
\usepackage{color}
\usepackage{url}
\usepackage{multirow}

\newtheorem{alg}{Algorithm}

\newsavebox{\savepar}

\newcommand{\bfx}{\mathbf {x}}
\newcommand{\bfy}{\mathbf {y}}
\newcommand{\bfz}{\mathbf {z}}
\newcommand{\bfb}{{\mathbf b}}

\newcommand{\bfg}{{\mathbf g}}

\newcommand{\bff}{{\mathbf f}}

\newcommand{\CP}{C_\mathrm{P}}
\newcommand{\CT}{C_\mathrm{T}}

\newcommand{\bitem}{\begin{enumerate}}
\newcommand{\eitem}{\end{enumerate}}

\begin{document}
%
%

%
%
%
%
%
%
%

\title{FDG-PET Parametric Imaging by Total Variation Minimization}
 \author[label1]{Hongbin~Guo\corref{cor1}}
 \cortext[cor1]{Corresponding author. Tel: 1-480-965-8002,   Fax: 1-480-965-4160.}
 \address[label1]{Arizona State University, Department of Mathematics and Statistics, Tempe, AZ  85287-1804. } 
 \ead{hguo1@asu.edu}
 \author[label1]{Rosemary~A~Renaut} 
 \author[label2]{Kewei~Chen}
 \address[label2]{Banner Alzheimer Institute and Banner Good
 Samaritan Positron Emission Tomography Center, Phoenix, AZ 85006}
 \author[label2]{Eric M Reiman}
\begin{frontmatter}
\begin{abstract}
Parametric imaging of the cerebral metabolic rate for glucose (CMRGlc) using [$^{18}$F]-fluorodeoxyglucose  positron emission tomography is considered.  Traditional imaging  is  hindered due to low signal to noise ratios  at individual voxels. We propose to  minimize the total variation of the tracer uptake rates  while requiring good fit of  traditional Patlak equations.   This minimization guarantees spatial homogeneity within brain regions and good distinction between brain regions. Brain phantom simulations demonstrate  significant improvement in quality of  images by the proposed method  as compared to Patlak images with post-filtering using Gaussian or median filters.

\end{abstract}


\end{frontmatter}


\section{Introduction}
We focus on  positron emission tomography (PET) parametric imaging
for estimating the cerebral metabolic rate of glucose (CMRGlc)  using the
[$^{18}$F]-fluorodeoxyglucose (FDG) tracer. The ability to derive
accurate parameters depends upon the quality of data, the quantification
method and the numerical algorithm. In this study, we refer to the time activity curve (TAC) from a given tissue location as the \textit{output}, tissue TAC or TTAC, and the TAC from  the blood pool (image-derived or arterial blood-sampled) as the \textit{input}, plasma TAC or PTAC.    Most existing quantification methods perform well for regions of interest
(ROIs), but are not good for voxel level quantification due to the high level of noise. These
include graphical methods, \cite{Patlak83,Logan90}, linear least
squares, the weighted integration method, \cite{Carson:98},
generalized linear least squares, \cite{Feng96,ChenGLLS:98},
nonlinear least squares (NLS) and weighted NLS.

All the algorithms listed above perform the quantification at each voxel location separately; they do not consider the kinetic similarities among neighboring  voxels within  functionally-defined regions. Thus, voxel-by-voxel variation in a
functionally-homogeneous region may be large because of  noise in the data. But, by incorporating the  spatial constraint that parameters in a functionally-homogeneous region should be similar, in any of the
above methods, the quality of the resulting  parametric image for the
CMRGlc may be improved.  Zhou et
al, \cite{Zhou2003}, for example, improved the parametric image quality by ridge
regression with constraints on the rate constants. There, the estimation of parameters uses a
linear components decomposition of the kinetics in which each component
represents a functional kinetic curve generated by clustering the TTACs, and the problem is solved voxel by voxel.
Here we propose the use of a total variation
(TV) penalty term which imposes spatial consistency between neighboring voxels.

The total variation penalty was first introduced in the context of image deblurring by Rudin  {\it et al}, 
\cite{Ruetal}.  TV can significantly
suppress noise while recovering  sharp edges because it does not
penalize  discontinuities. It has received much theoretical
research attention and been utilized in many signal
and image processing applications. While it was introduced for PET image
reconstruction by Jonsson {\it et al}, \cite{Jonsson98}, and Kisilev
{\it et al}, \cite{KisilevZZ01},  it has apparently not been  applied for  parametric PET imaging. Instead of calculating the  uptake rate
for each voxel by  Patlak's method, we propose to minimize the TV of the uptake rate over the entire  image while also maintaining  a good least squares fit for the
Patlak equations at all voxels. Thus the parameters of the whole
image are spatially related by the TV and solved simultaneously. The
resultant parametric image is expected to have spatial homogeneity
over brain regions with similar kinetics and distinct edges between brain
regions that have different kinetics. This is  validated by phantom simulations.

In addition to proposing the new model with the TV penalty, we also pay careful attention to the computational complexity of the algorithm by taking advantage of
implementations for large scale sparse matrix computations.
In contrast to approximating the  Hessian matrix, as is typical for
quasi-Newton methods, our algorithm  explicitly and accurately
calculates both gradient and Hessian terms. The Hessian is
efficiently recalculated at  each iteration because of its sparsity.
The Quasi-Minimal Residual (QMR) method, \cite{Freund91}, is used to solve the
resulting  large-scale linear systems. With this efficient implementation, the procedure described here is computationally feasible.

The rest of the paper is organized as follows: The new algorithm with TV penalty is introduced in section~\ref{sec:method}, with relevant computational issues detailed in the appendix. The  experimental data sets are described in  section~\ref{sec:valid} and  
results  reported in section~\ref{sec:results}. Issues relevant to the proposed TV-Patlak method and computational aspects are discussed in section~\ref{sec:dis}.  Conclusions are presented in section~\ref{sec:conc}. 

\section{Methods}\label{sec:method}
The Patlak plot has been developed for systems with irreversible trapping \cite{Patlak83}. Most often it is applied for the analysis of FDG. 
The 
measured TTAC undergoes a transformation and is plotted against a \textit{normalized time}. It is given by the expression
 \begin{equation}\label{eq:patlak}
 \frac{\CT(t)}{\CP(t)}\approx K \frac{\int_0^t\CP(s){\rm d}s}{\CP(t)}+V,
 \end{equation} 
where $\CT(t)$ is the measured TTAC, (in counts/min/g) and $\CP$ is the PTAC (in counts/min/ml), i.e. the FDG concentration in plasma.
For systems with irreversible compartments this plot yields a  straight line after sufficient equilibration time. 
For the FDG tracer, the slope  $K$ represents the uptake rate which, together with lumped constant (LC) and glucose concentration in plasma ($C_{\mathrm{pg}}$) allows easy calculation of the CMRGlc=$KC_{\mathrm{pg}}$/LC, (in mg/min/$100$g). The intercept $V$ is given by  $V_0+vB$ where $V_0$ is the distribution volume  of the reversible compartment and $vB$ is the fractional blood volume.

The linear relationship  (\ref{eq:patlak}) can be rewritten as
 \begin{equation}\label{eq:patlak_lin}
 \Big(\int_0^t\CP(s){\rm d}s\Big) K+\CP(t)V\approx \CT(t),
 \end{equation}
and, assuming $m$ dynamic frames over the period of equilibration, its discretized version  is
\begin{equation}\label{eq:patlakvec_des}
\Big(\int_0^{t_j}\CP(s){\rm d}s\Big)K +\CP(t_j)V\approx \CT(t_j),
\ \ j=1,\cdots,m.
\end{equation}
In matrix format,
\begin{equation}\label{eq:patlakvec}
A\left(\begin{array}{c}
          K \\
          V
        \end{array} \right)\approx \bfb,
\end{equation}
where $A$ is a $m$-by-$2$ matrix,
$$A=\left(\begin{array}{cc}
          \int_0^{t_1}C_\mathrm{P}(s){\rm d}s,&C_\mathrm{P}(t_1) \\
          \int_0^{t_2}C_\mathrm{P}(s){\rm d}s,&C_\mathrm{P}(t_2) \\
          \vdots&\vdots\\
          \int_0^{t_m}C_\mathrm{P}(s){\rm d}s,&C_\mathrm{P}(t_m)
        \end{array} \right),$$
and vector $\bfb=(\CT(t_1), \CT(t_2), \cdots, \CT(t_m))^T$. If we were to solve (\ref{eq:patlakvec}) for each voxel independently  we would obtain a parametric image  lacking spatial homogeneity and with low signal-to-noise ratio (SNR). Image denoising techniques could then be applied as a separate task to improve the image quality. Instead, obtaining all voxel parameters as a result of a global optimization algorithm with a TV penalty for the entire image, the necessity for postprocessing should be eliminated.

Limiting the discussion here to $2$D images (although our application of TV is $3$D), we select the active voxels to be quantified by the application of a brain mask, yielding a total of $N$ voxels.   Equation (\ref{eq:patlakvec}) holds with common matrix $A$ dependent on  $\CP(t)$  for each voxel $i$, but with $K$, $V$ and $\bfb$  replaced by $K^{(i)}$, $V^{(i)}$, and $\bfb_i$, respectively, where $\bfb_i$ is obtained from the TTAC for voxel $i$.  Collecting the unknowns of these $N$ voxels in  vectors of uptake rates and intercepts  
$$\bfx=(K^{(1)}, K^{(2)}, \cdots, K^{(N)})^T \quad \mathrm{and} \quad \bfy=(V^{(1)}, V^{(2)}, \cdots, V^{(N)})^T,$$
and requiring (\ref{eq:patlakvec}) in the least squares sense over all voxels, while maintaining minimal TV of the uptake rate  for the selected image voxels, yields the global minimization problem
\begin{eqnarray}
\label{tv-patlak}
  ({\rm TV-Patlak}):&& \quad \min \Phi(\bfx;\bfy),\\
\Phi(\bfx;\bfy)&=&  \|\bfx\|_{\mathrm{TV},\beta}+ \alpha \sum_{i=1}^N \|W_i(A (x_i,y_i)^T-\bfb_i) \|_2^2. \label{eq:TVfunct}
\end{eqnarray}
Here the total variation norm is given by $\|\bfx\|_{\mathrm{TV},\beta}=\sum_{i=1}^N\phi_i(\bfx)$ with
$\phi_i(\bfx)=\sqrt{(x_i-x_{i_r})^2+(x_i-x_{i_b})^2+\beta^2}$, and
 $x_{i_r}$ and $ x_{i_b}$ are the values  associated with
voxels to the right and below  voxel $i$. Theoretically the TV
norm is $\|\bfx\|_{\mathrm{TV},0}$, which is a seminorm on a space of  bounded
variation, \cite{Vogelbook02}. The small constant $\beta$ is
used to avoid the numerical difficulty due to the lack of differentiability at the origin of $\phi_i$ for $\beta=0$. 
The diagonal weight matrix for voxel $i$ is given by
\begin{equation}\label{eq:weiht}
W_i={\rm diag}\left(\sqrt{\frac{\Delta t_j}{b_{ij}e^{\lambda t_j}}} \right),\quad j=1,\cdots,m,
\end{equation}
where $\Delta t_j$ is the scan duration of the frame at time $t_j$, $\lambda$ is the tracer's decay constant and $b_{ij}$ is the value of the $i^{\mathrm{th}}$ TTAC at frame $j$. This weighting  is consistent with using a simulation with variance   
\begin{equation}\label{eq:var_y}
\mathrm {Var}(\CT(t_j))=Sc \frac{\CT(t_j) e^{\lambda t_j}}{\Delta t_j},
\end{equation}
\cite{logan2001a}. 
$Sc$ is a common scale factor that need not be made explicit here because it  is absorbed into the parameter $\alpha$ in (\ref{eq:TVfunct}).

The objective function in (\ref{tv-patlak}) is  convex,  and can  be solved using a standard Newton-type algorithm, \cite{Vogelbook02} Chapter 8. To simplify the  expressions we introduce the vector $\bfz=[\bfx;\bfy]$.
\begin{alg} \label{alg:tikTV}
Given initial guess  $\bfz=[\bfx;\bfy]$ and  tolerance $\epsilon > 0$\\
{\bf Repeat}
\begin{enumerate}
\item \textbf{Solve} for  $\Delta\bfz$ in
 \begin{equation}\label{stepupdate}
\nabla^2 \Phi(\bfz) \Delta\bfz=-\nabla \Phi(\bfz).
\end{equation}
\item \textbf{Stop} if $|\nabla \Phi^T\Delta \bfz| \le \epsilon.$
\item \textbf{Line search:}  Choose step size $s$.
\item\textbf{Update:} $\bfz= \bfz + s\Delta\bfz$.
\end{enumerate}
\end{alg}
Further details on the calculation of gradient vector $\nabla \Phi(\bfz)$ and  Hessian matrix $\nabla^2 \Phi(\bfz)$  are provided in  the appendix. Some other  aspects of the algorithm, including discussion about constants $\alpha$ and $\beta$, here chosen to be $0.2$ and $10^{-8}$, respectively,  as well as  other approaches for the solution of the  TV problem are discussed in section~\ref{sec:dis}.

\section{Experimental Data } \label{sec:valid}
To validate the proposed parametric imaging method we performed
experiments with simulated  data.
The MRI-based high-resolution Zubal head phantom is used to define the brain structures, \cite{Zubalphantom}. Each voxel in the slice of $256\times256$ voxels is of size $1.5\times1.5$mm$^2$. There are  $128$ slices for the entire head and $62$ defined anatomical, neurological, and taxonomical structures. The $11$ regions for slice $64$, representing a total $13708$ voxels, are given in table~\ref{tab:simki}. For the purposes of the simulations, well-accepted values of  the kinetic rate parameters of these structures for the two-tissue compartmental model of the FDG tracer \cite{Huetal:80}, are assigned. Notice that in order to better model the true biochemical
process we assume $k_4>0$. Because $k_4$ is, however, relatively small,  Patlak's
method, for which it is assumed $k_4=0$, is still a suitable graphical method for quantifying the uptake rate.

\begin{table}[htbp]
\caption{Brain regions and  rate constants\cite{Huetal:80} for slice $64$ of the Zubal head phantom \cite{Zubalphantom}.}
\label{tab:simki}

\begin{center}
\begin{tabular}{|l|c|c|c|c|}
\hline
Brain Regions & $K_1$ &$k_2$ &$k_3$&$k_4$\\
&ml/min/g &1/min&1/min&1/min\\
\hline
 frontal lobes&&&&\\
 occipital lobes &&&&\\
 insula cortex&  $0.102$ & $0.130$ & $0.062$&$0.0068$\\
 temporal lobes&&&&\\
 globus pallidus&&&&\\ \hline
thalamus &$0.082$& $0.105$& $0.060$&$0.0068$ \\ \hline
 putamen&  $0.070$&$0.070$& $0.054$&$0.0068$\\
 caudate nucleus&&&&\\ \hline
 internal capsule &&&& \\
 corpus collosum &  $0.054$& $0.109$& $0.045$&$0.0058$\\
other  white matter&&&& \\
\hline
\end{tabular}
   \end{center}
\end{table}

The noise-free input function, with values given in kBq/ml, is given by the formulation introduced
in \cite{Guoinpf07},
\begin{equation}\label{inpf.eq}
u(t)=\left\{\begin{array}{ll}
0 & t\in[0,0.25]\\
339.03(t-0.25)& t \in [0.25, 0.4433]\\  
-214.656t+  160.691 & t \in [0.4433, 0.65]\\
21.165 e^{-0.7501(t-0.65)^{0.2359}}  & t >0.65.
\end{array}\right.
\end{equation}
Although any reasonable input function, including clinical plasma samples, could be used for the simulation this formulation was validated as providing a good approximation to the plasma samples of a healthy subject.
Given the input function, the \textbf{exact} phantom can then be generated using the rate constants for the structures detailed in table~\ref{tab:simki}, \cite{Sokoloffs}. The output time frames were generated assuming time frames with durations $\Delta t_j$, $j=1,\cdots,m$, $m=22 $, given in minutes,  $0.2$,
$8\times 0.0333$, $2\times 0.1667$, $0.2$, $0.5$, $2\times 1$,
$2\times 1.5$,  $3.5$,  $2\times 5$,  $10$ and $30$.



In our experiments, in order to control the computational overhead of the reconstructions, we reduced the image size of the phantom from
$256\times 256$ to $128\times 128$. To do this we needed to relabel the structure assigned to a given voxel. This was done by averaging the quantity $K_1k_3/(k_2+k_3)$ over the $2$-by-$2$ neighboring voxels   at the finer resolution. Then the voxel at the coarser level was labeled as belonging to the structure for which this average is closest to the structure value.   The kinetic values were then assigned, the TTAC output values calculated, and then projected to the sinogram space yielding projected noise-free sinogram data $P$.  For simplification,  the instrumental and physical effects,
including attenuation, Compton scattering, decay and random
coincidences, were not simulated. 
Poisson noise was then added to the projected data using $S=\mathtt{poissrnd}(P)$ where  the Matlab{\textregistered,\cite{matlab}} function \texttt{poissrnd} uses vector $P$ as  the means of Poisson densities to  generate the noisy sinogram data $S$. Based on this noisy sinogram data, concentration  images are reconstructed using the Expectation-Maximization (EM) algorithm, \cite{Kaufman1987}.

Several data sets were generated: To investigate errors introduced
by violation of the irreversibility assumption, $k_4=0$, we tested both $k_4=0$ and
$k_4>0$. Similarly, we tested both with and without noise in the
sinograms to investigate the effects of the proposed method due to
noise in the sinograms.  Gaussian noise with noise levels  $0\%$, $5\%$, $10\%$, $15\%$ and $20\%$ was added to the input function, i.e.
\begin{equation}\label{eq:noisyinput}
 \CP(t_j)=u(t_j)(1+\mathrm{CV}\eta_j),
\end{equation}
where $\eta_j$ is selected from a standard normal distribution (G$(0,1)$), and $CV=0$, $0.05$, $0.10$, $0.15$ and $0.20$.  $100$ random realizations are tested in each case.

We summarize  the test data sets  in table
\ref{tab:test-cases}, using a character triple to classify each test. The first character of the triple indicates whether $k_4=0$ or $k_4>0$, $0$ or $+$, respectively. The second character indicates the noise level on the input function, $0$ or $+$, for noise-free, or with added noise, respectively. To indicate the noise level $+$ is replaced by  $1$, $2$, $3$ and $4$ to indicate
noise levels $5\%$, $10\%$, $15\%$ and $20\%$  as necessary. The third character indicates whether noise is added to the sinogram, again $0$ or $+$, respectively. Therefore, for example, $+3+$ represents the
test case for  $k_4>0$, $15\%$ noise in $u(t)$ and Poisson
noise added to the sinogram.

\begin{table}[htbp]
\begin{center}
\caption{Summary of the test cases. Here $0$ in columns two and three indicates the noise-free case, while $+$ indicates noise was added. In column one the $0$ indicates $k_4=0$. The same comments apply, but with the irreversibility assumption  violated for all triples starting with $+$, indicating $k_4>0$. \label{tab:test-cases}}

\begin{tabular}{|c|c|c|l|}
\hline
$k_4$ & $u(t)$ & Sinogram & Comments \\ \hline
  & $0$ & $0$ &  Errors only caused by reconstruction \\ \cline{2-4}
$0$ & $0$ & $+$ & Errors in TTACs caused by reconstruction plus Poisson
noise \\ \cline{2-4}
  & $+$ & $0$ & Errors in $\CP(t)$ \\ \cline{2-4}
  & $+$ & $+$ & General case for irreversible compartmental model \\ \hline
\end{tabular}
   \end{center}
\end{table}

\section{Results}\label{sec:results}
To evaluate the simulations quantitatively  we define the relative error of the $k^{\mathrm{th}}$ realization for voxel $i$:
$$r_{ik}=\frac{\widehat{K^{(i)}_k}-K_{\mathrm{true}}^{(i)}}{K_{\mathrm{true}}^{(i)}},\quad i=1,\cdots,N,\ k=1,\cdots,100,$$
where $\widehat{K^{(i)}_k}$ is the estimated value of the true value of the uptake rate $K_{\mathrm{true}}^{(i)}$ at the  $k^{\mathrm{th}}$ realization for voxel $i$. Over $100$ random simulations,  and $N$  voxels,  we calculate the bias, i.e. the mean of the relative errors, and the deviation from the mean:
\begin{equation}\label{eq:relerr}
\bar{r}=\frac{\sum_{i=1}^N\sum_{k=1}^{100}r_{ik}}{100N},\quad d=\frac{\sum_{i=1}^N\sum_{k=1}^{100}|r_{ik}-\bar{r}|}{100N}.
\end{equation}
Absolute  relative errors $|r_{ik}|$ and associated mean
$\bar{R}={\sum_{i=1}^N\sum_{k=1}^{100}|r_{ik}|}/{100N}$, and
deviation,
$D={\sum_{i=1}^N\sum_{k=1}^{100}||r_{ik}|-\bar{R}|}/{N}$, are
also calculated. Note the deviations are  $l_1$ measurements which do not overweight outlier and large error samples, as is the
case for the $l_2$-based measurements such as the root mean squared
error. 

\begin{figure}[htpb]
\centerline{
\includegraphics[height=1.5in,width=1.7in]{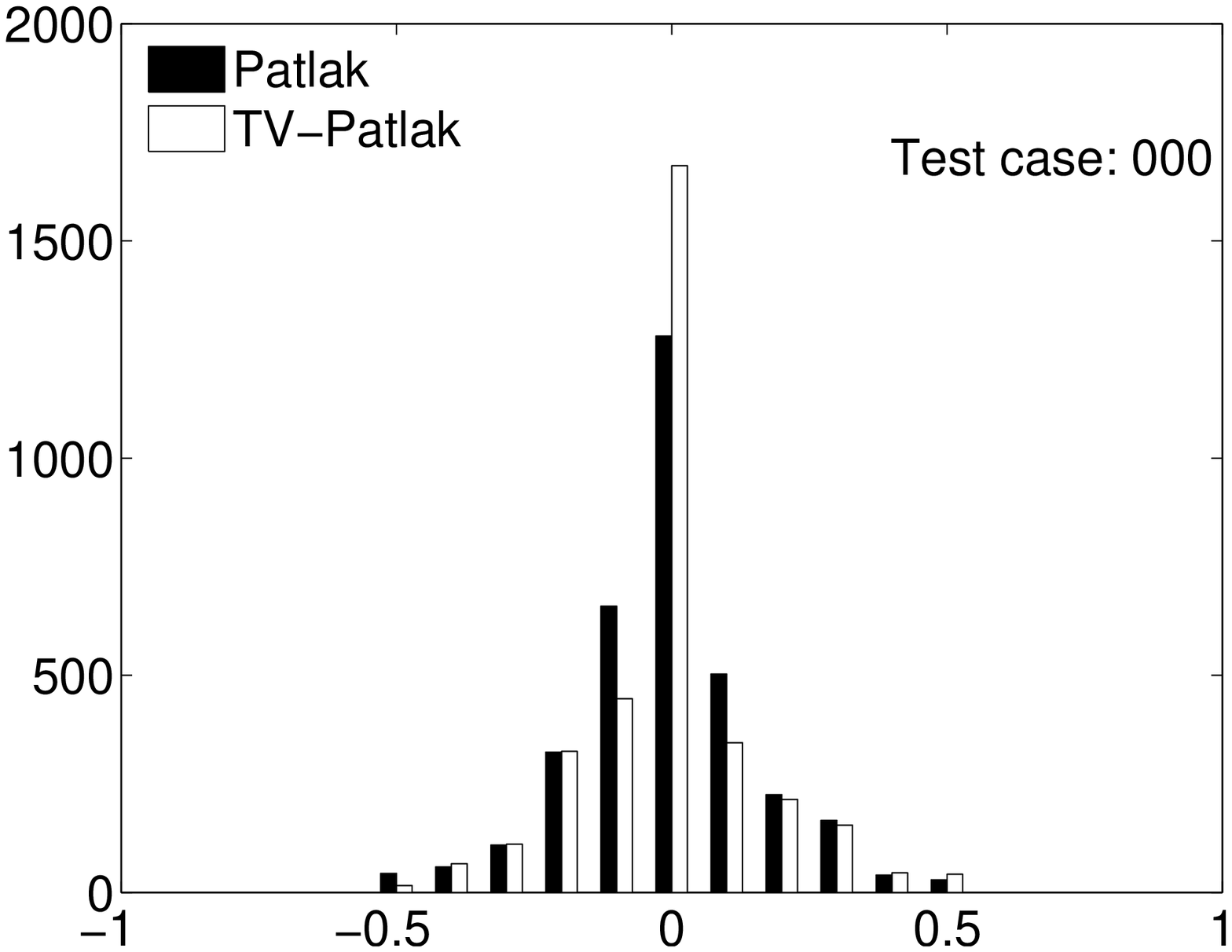}
\includegraphics[height=1.5in,width=3in]{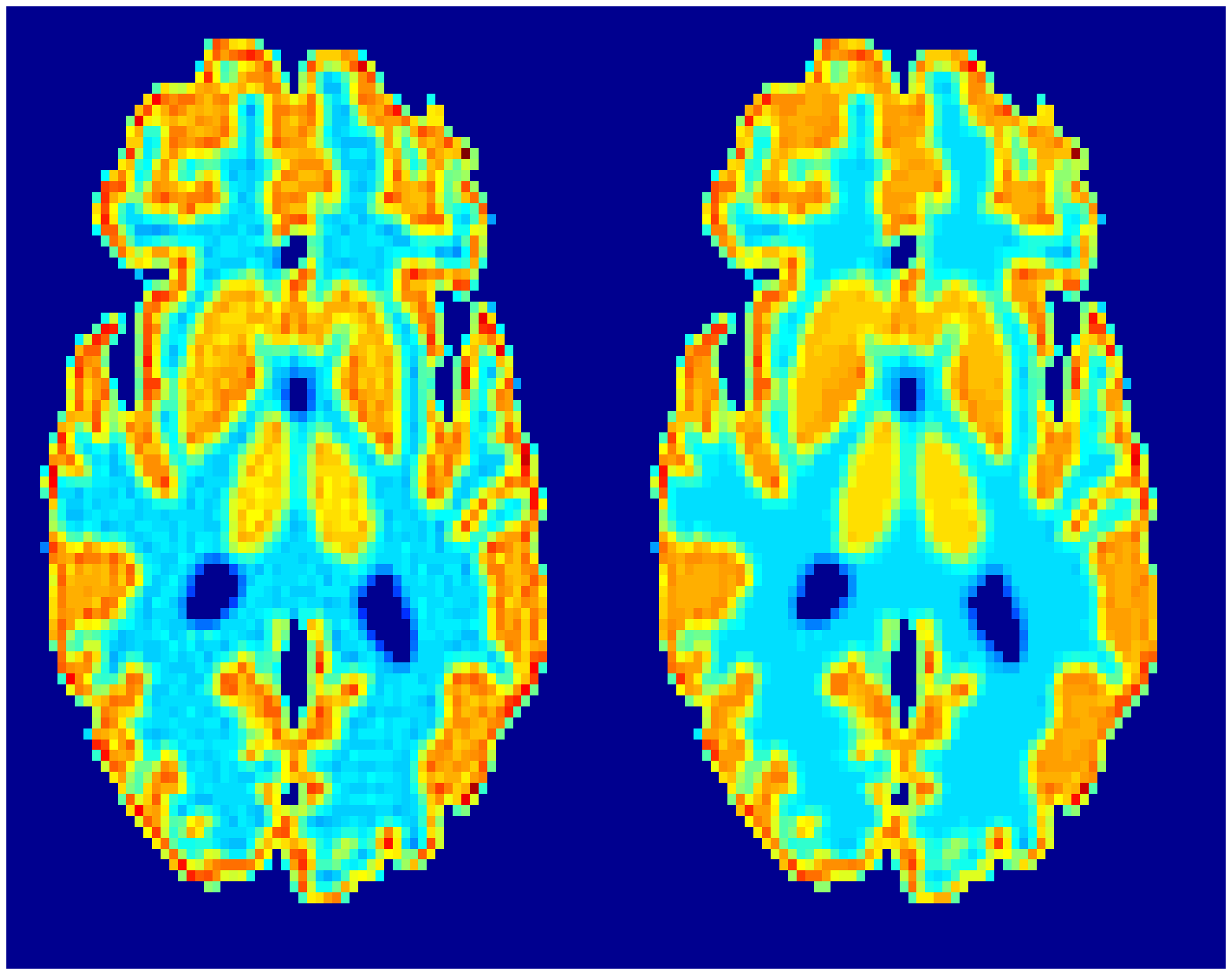}
}
\centerline{
\includegraphics[height=1.5in,width=1.7in]{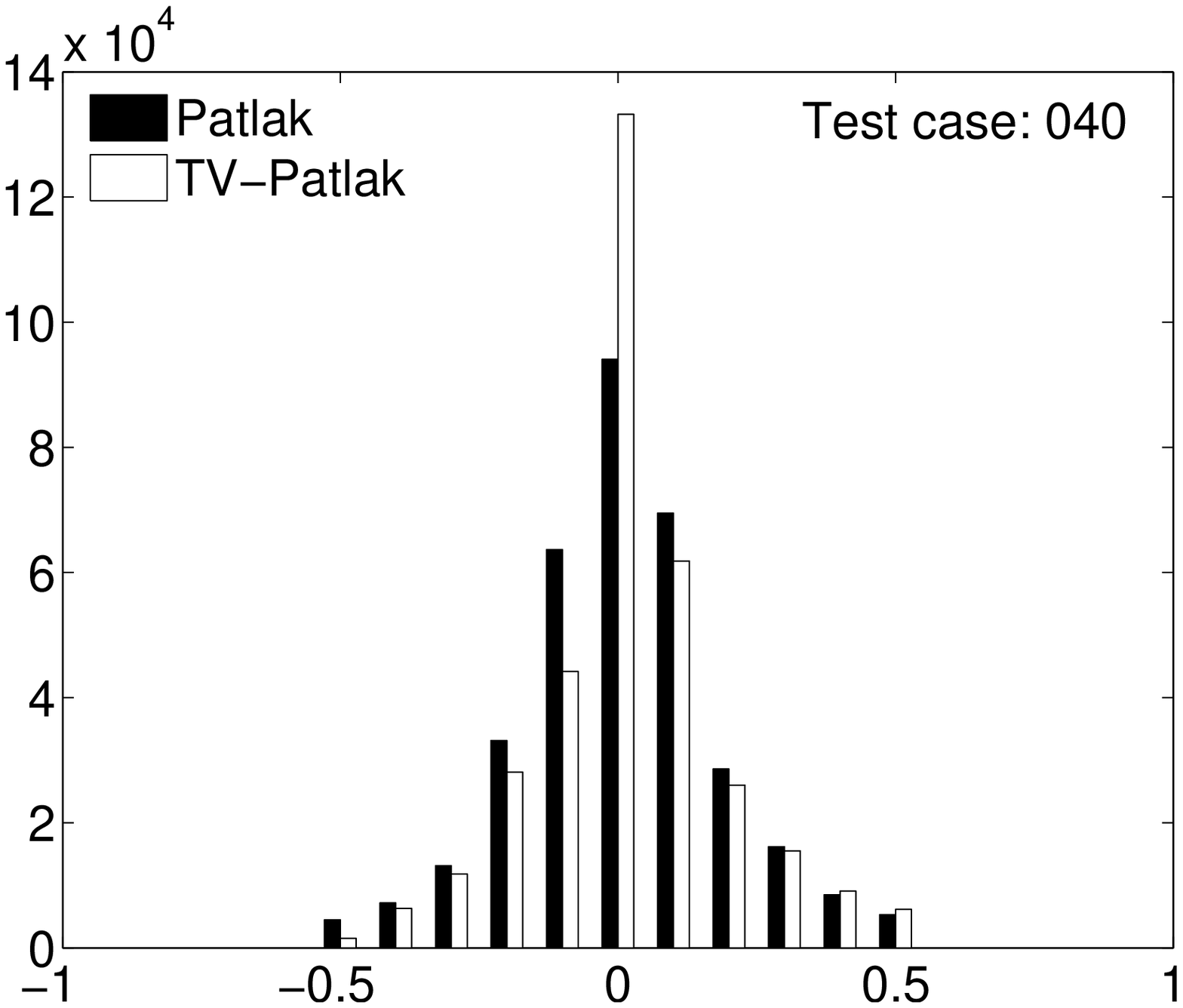}
\includegraphics[height=1.5in,width=3in]{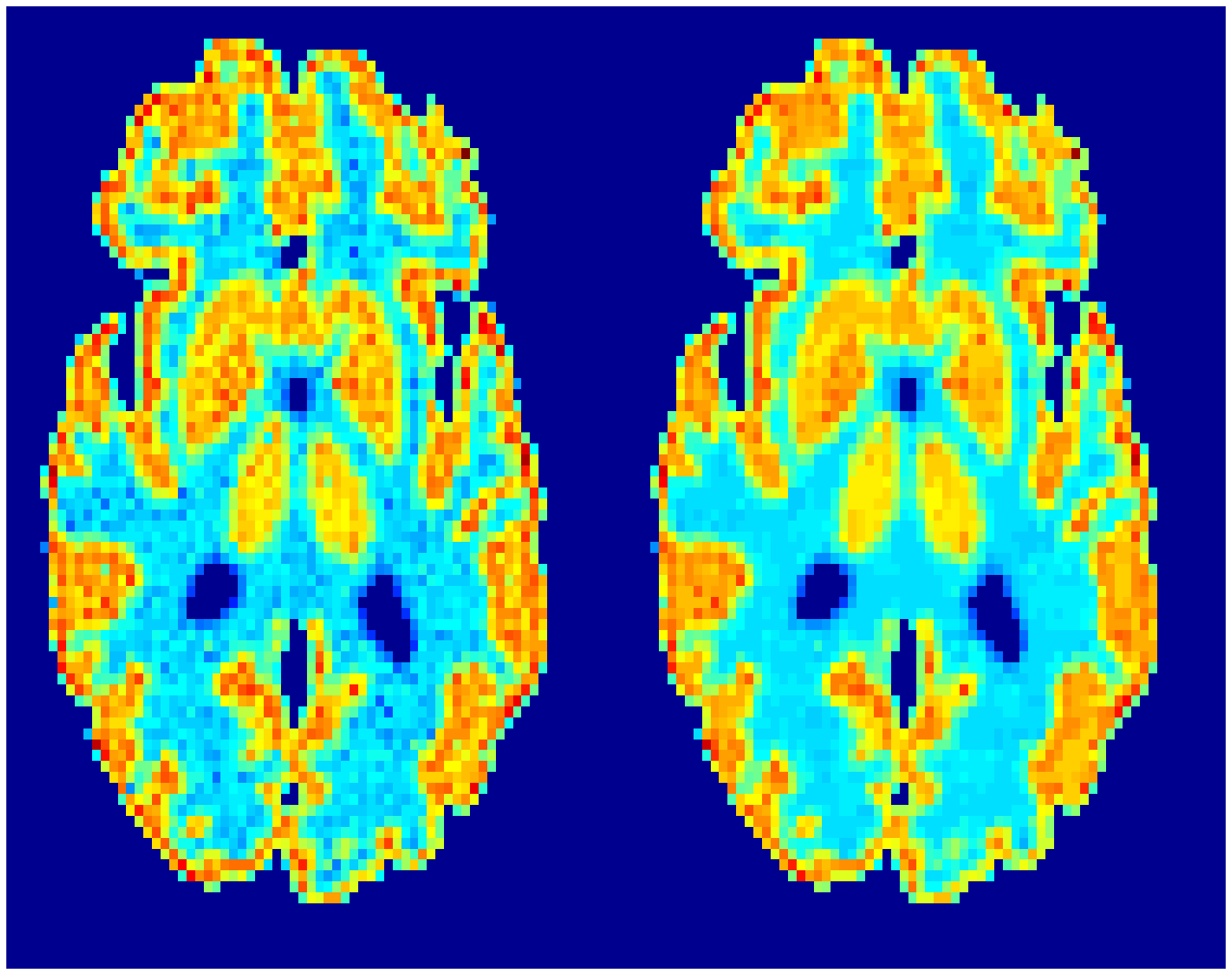}
}
\centerline{
\includegraphics[height=1.5in,width=1.7in]{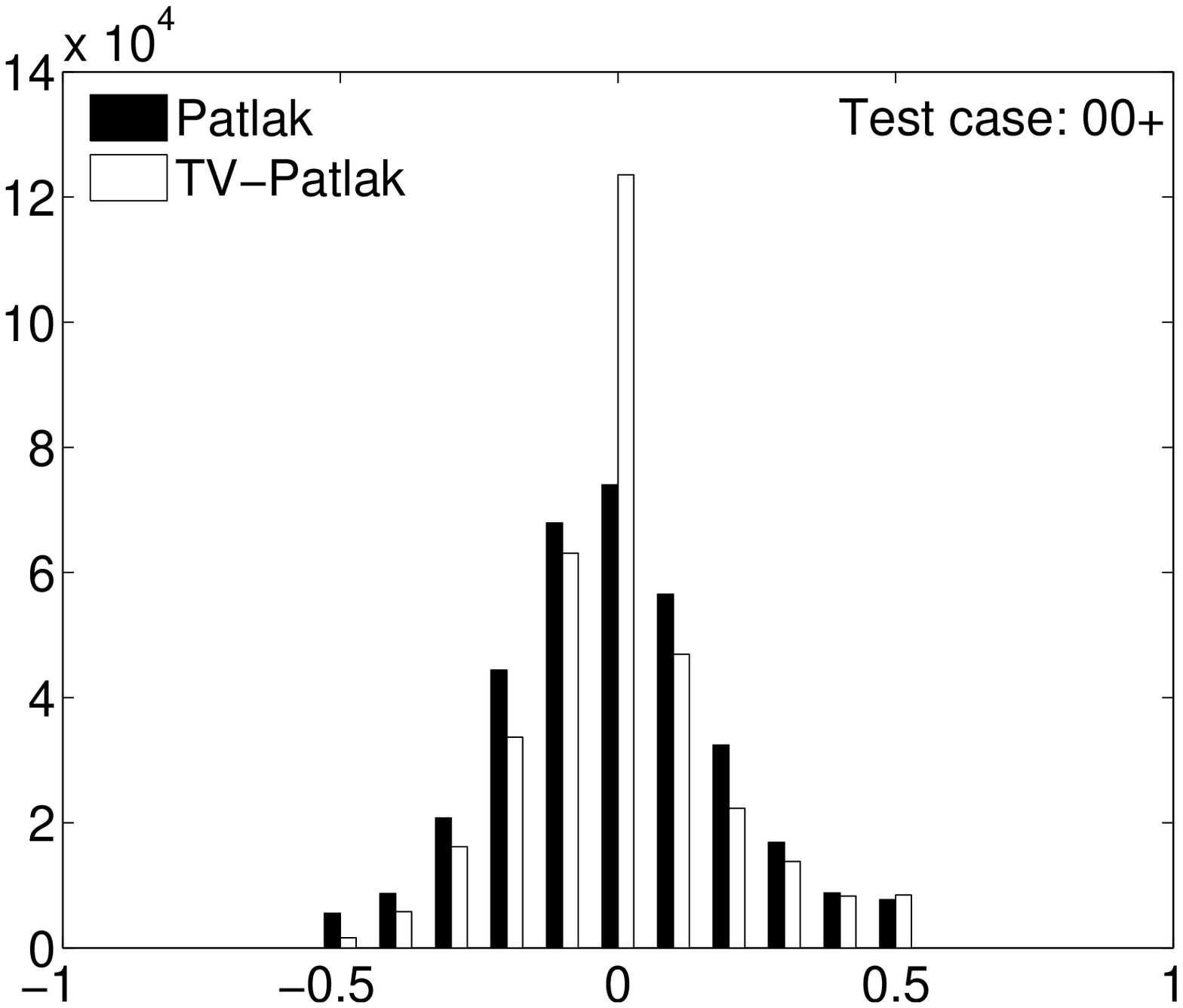}
\includegraphics[height=1.5in,width=3in]{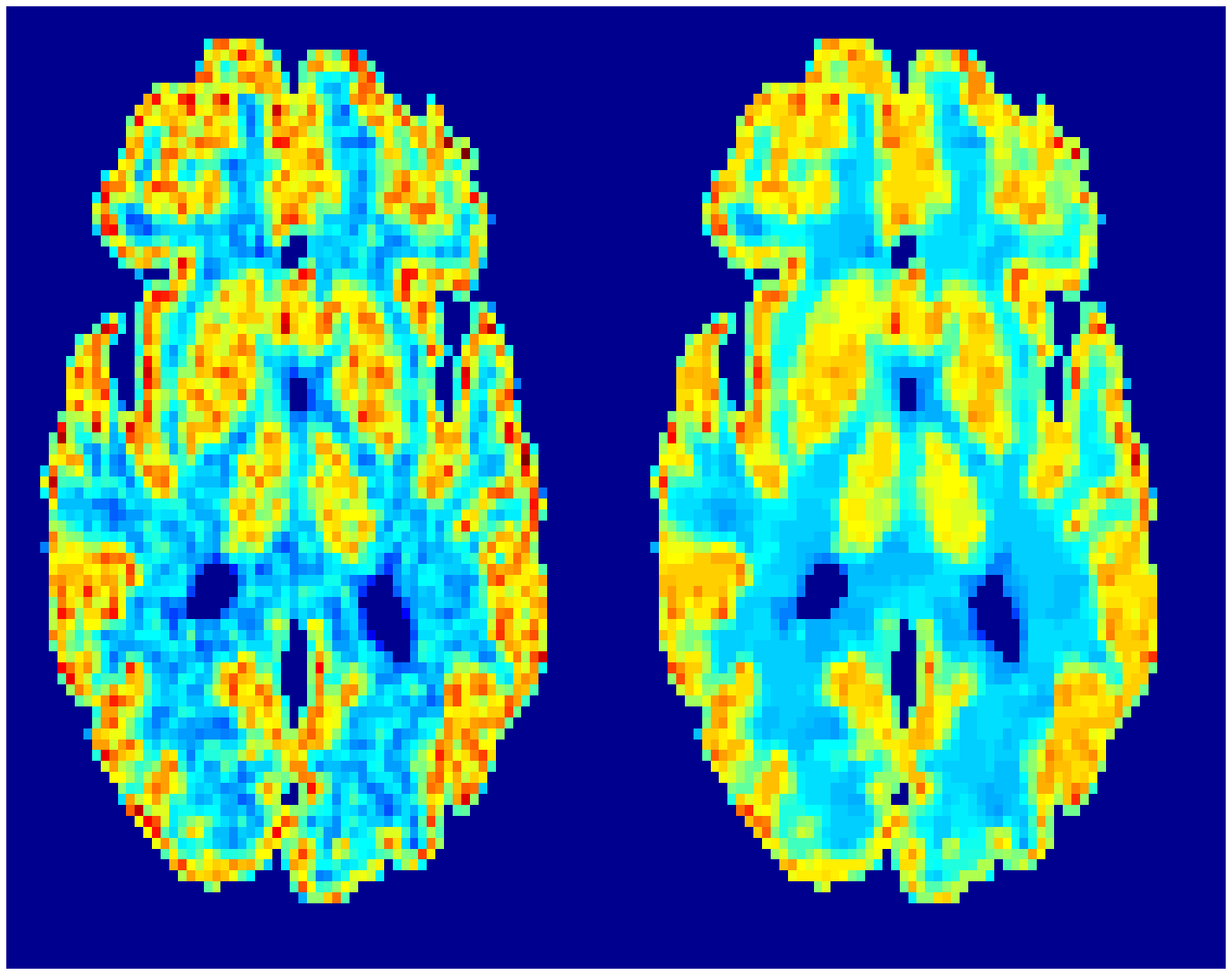}
}
\centerline{
\includegraphics[height=1.5in,width=1.7in]{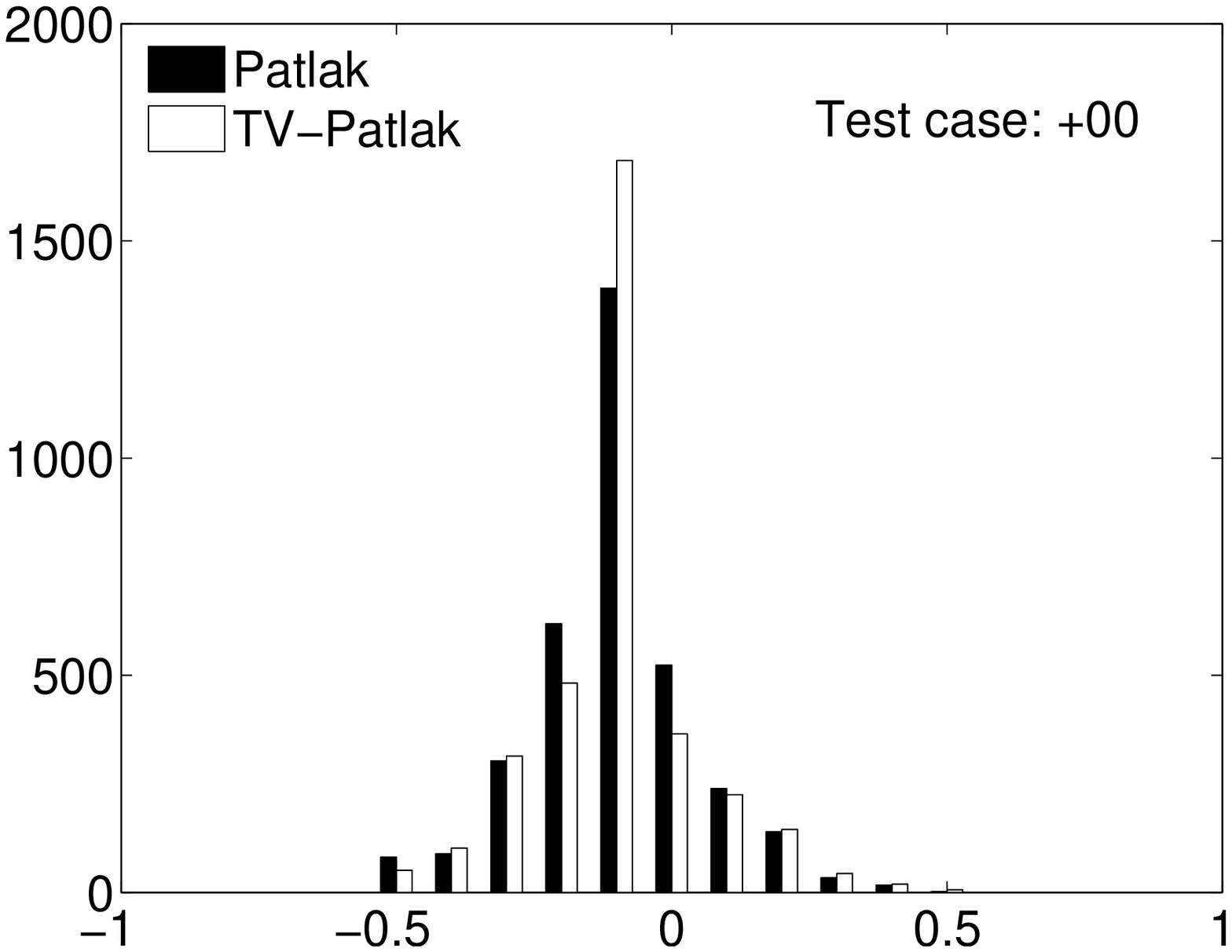}
\includegraphics[height=1.5in,width=3in]{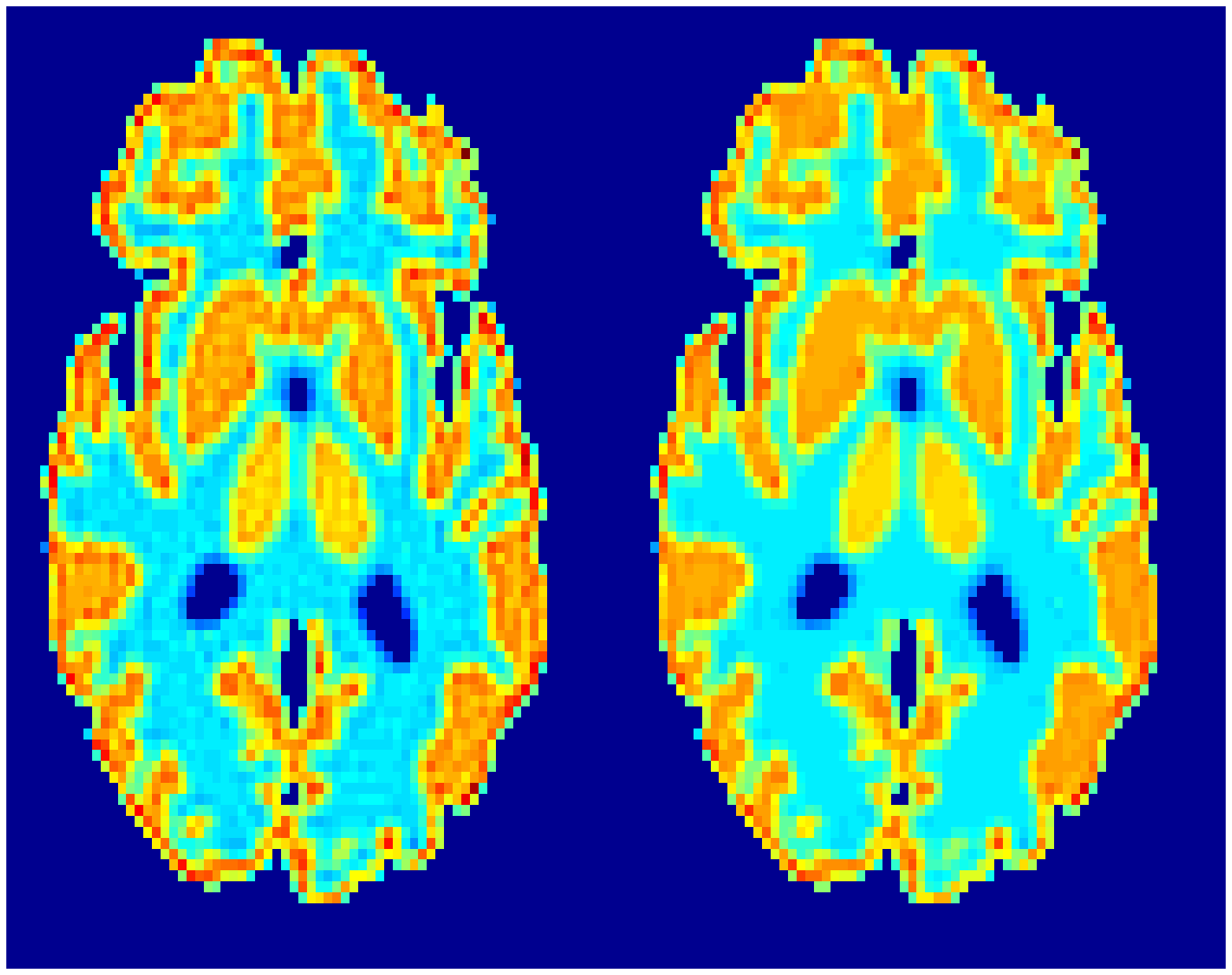}
}
\caption{Comparison of Patlak and TV-Patlak for imaging the uptake rate $K$. In each case on the left is the histogram for the relative error as compared to the exact simulated value, in the middle the Patlak image and on the right the TV-Patlak image. The first row compares the estimation by Patlak and TV-Patlak with no noise added. The second, third and fourth rows provide the comparison with $20\%$ noise added in the input, Poisson noise added to the sinogram, and positive $k_4$, respectively.\label{fig:err-single}}
\end{figure}

\begin{figure}[bp]
\centerline{
\includegraphics[height=1.5in,width=1.7in]{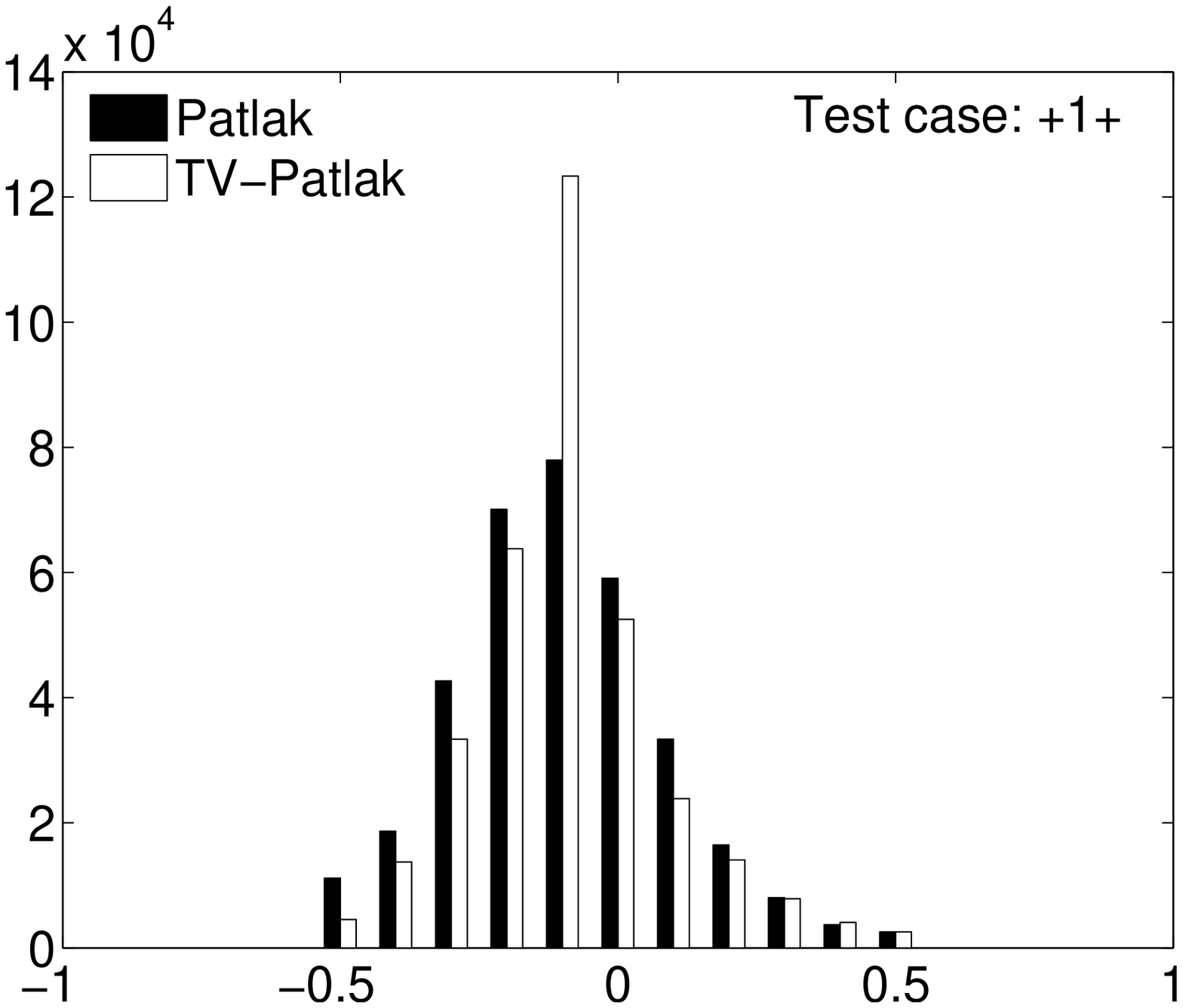}
\includegraphics[height=1.5in,width=3in]{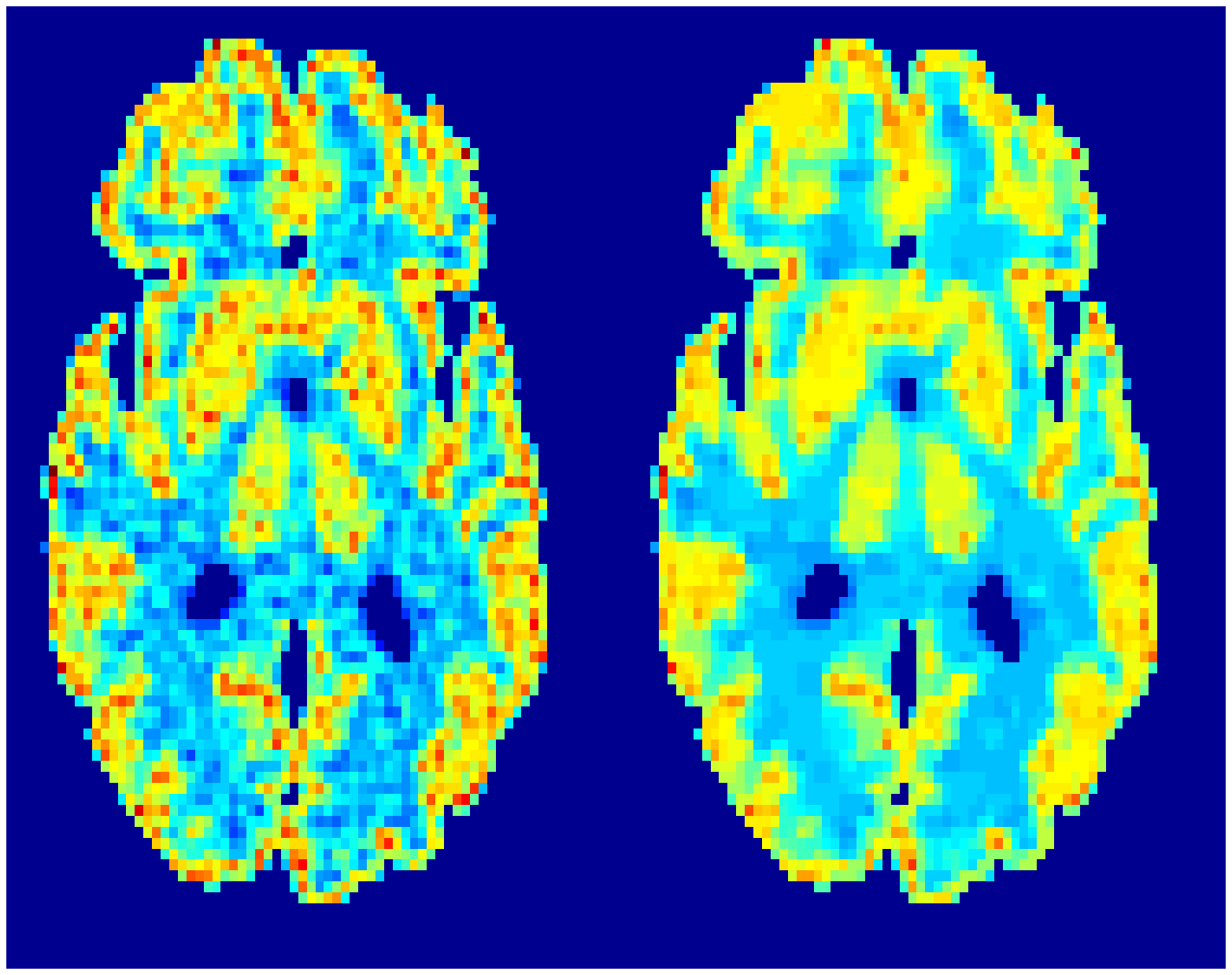}
}
\centerline{
\includegraphics[height=1.5in,width=1.7in]{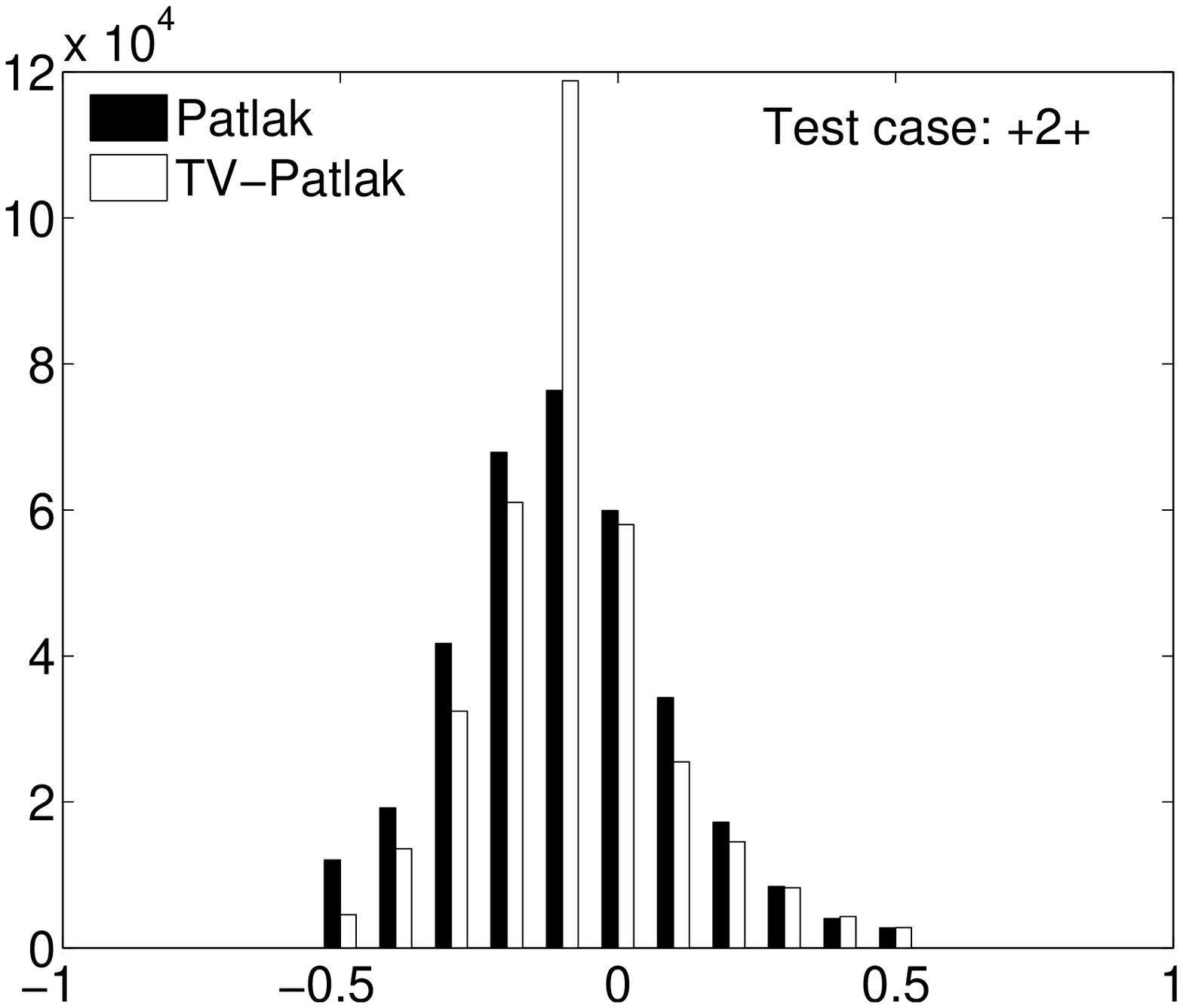}
\includegraphics[height=1.5in,width=3in]{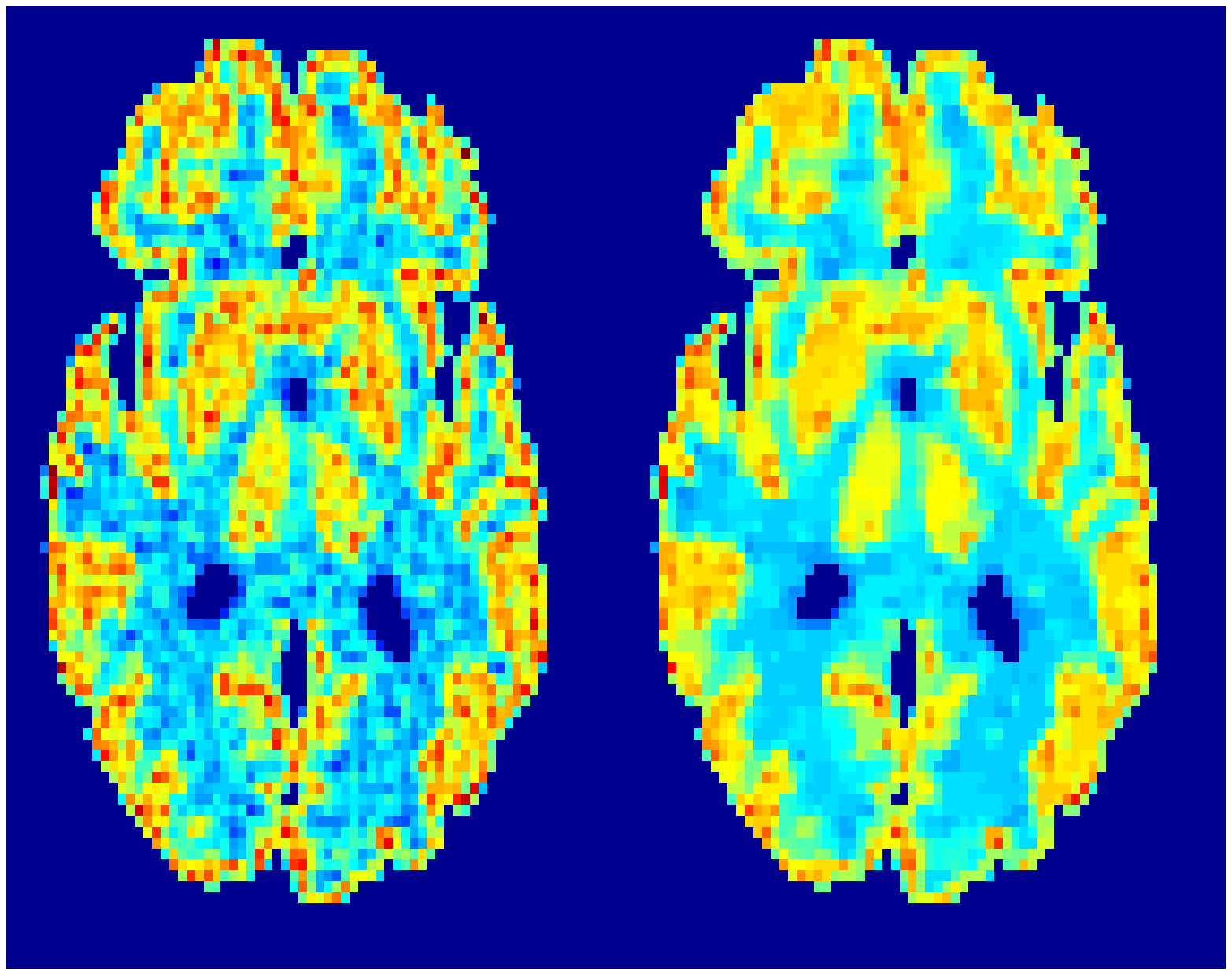}
}
\centerline{
\includegraphics[height=1.5in,width=1.7in]{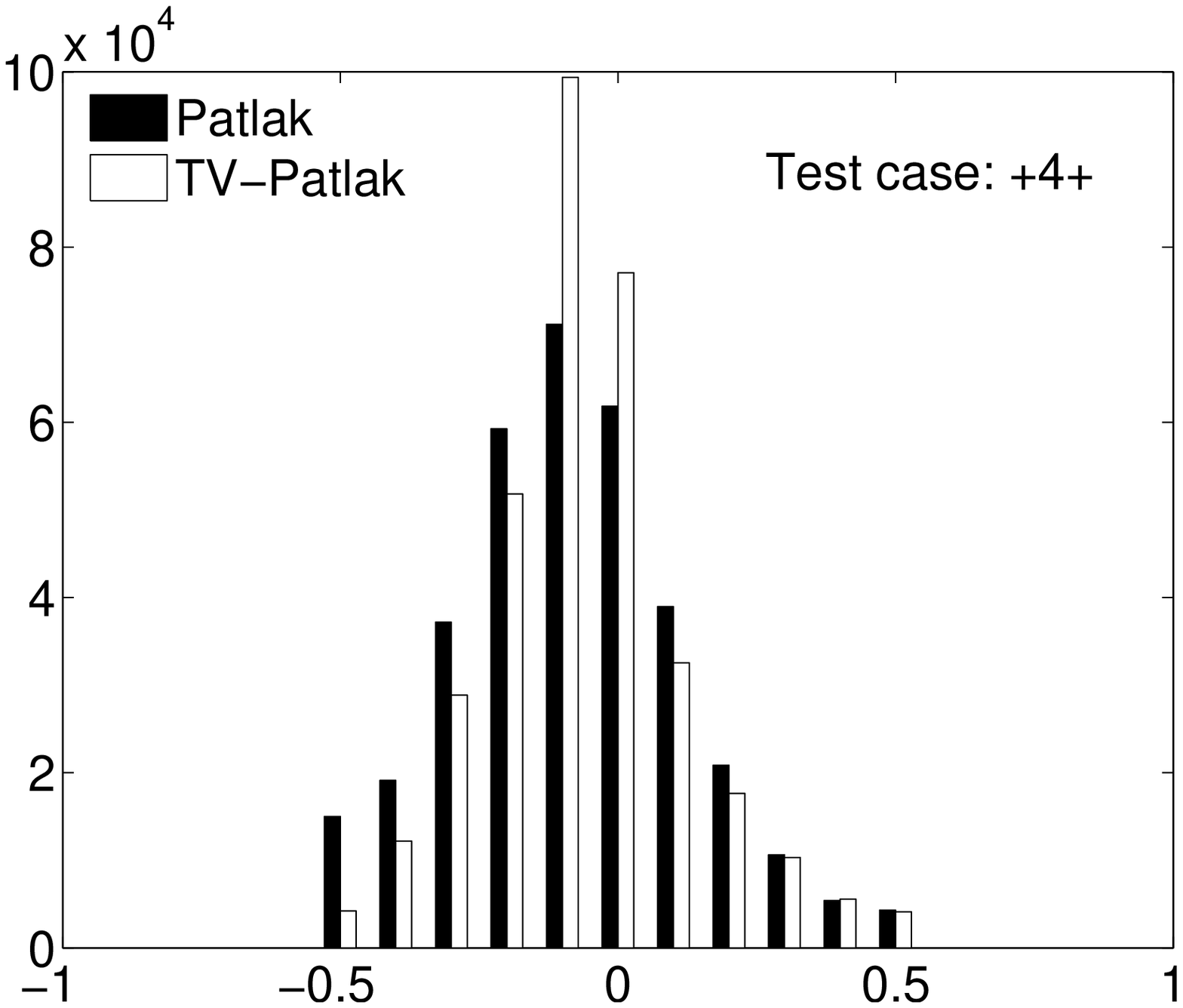}
\includegraphics[height=1.5in,width=3in]{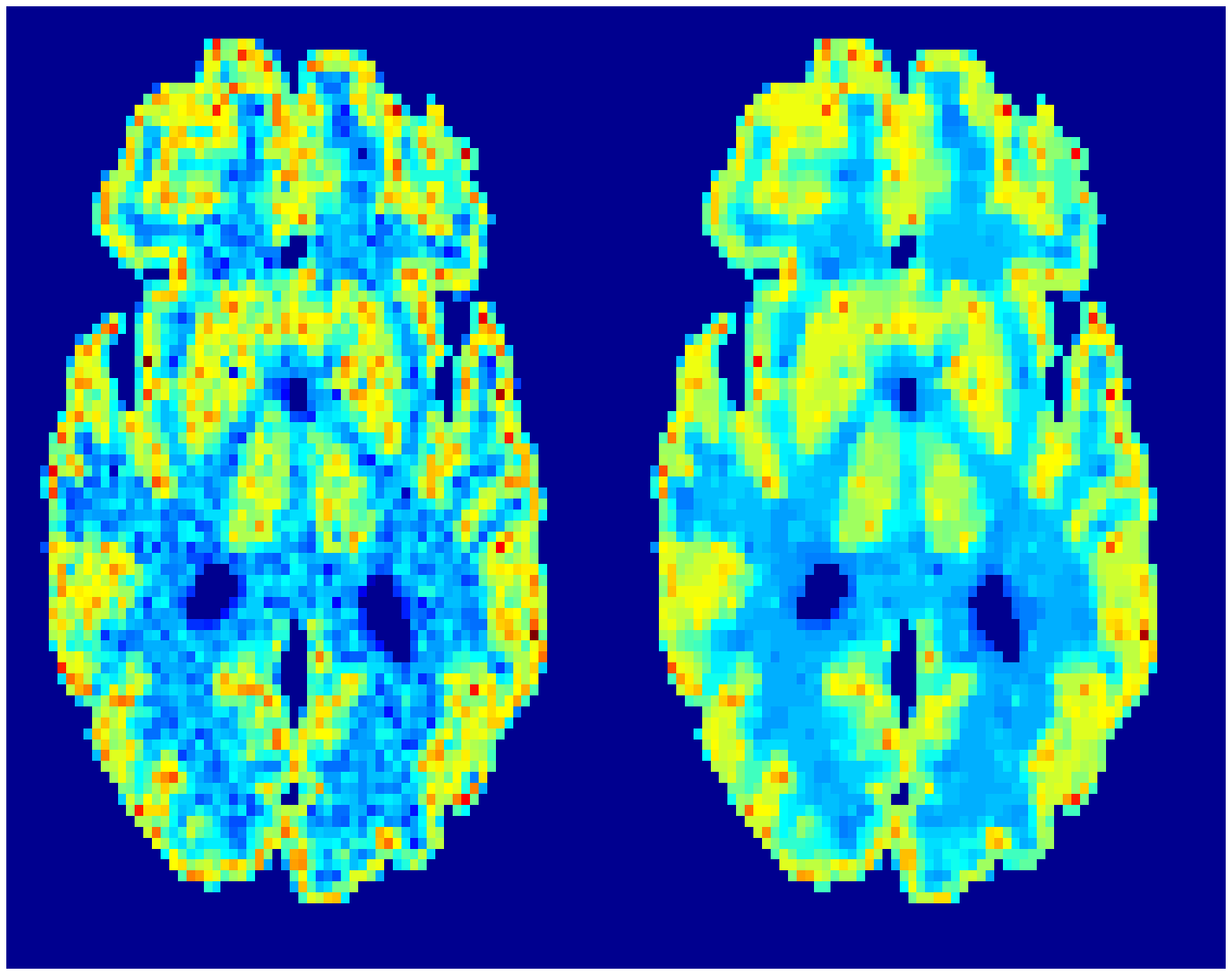}
} \caption{Comparison of Patlak and TV-Patlak for imaging the uptake rate $K$.  In each case on the left is the histogram for the relative error as compared to the exact simulated value, in the middle the Patlak image and on the right the TV-Patlak image. In each case Poisson noise is added to the sinogram and $k_4>0$. The level of noise for $100$ realizations added to the input is $5\%$, $10\%$ and $20\%$ respectively, for each row from top to bottom.\label{fig:err-general}}
\end{figure}

In the images shown in the figures we illustrate the calculated uptake rates $K$ of the FDG. Images for the CMRGlc can be obtained by directly scaling $K$. In figure~\ref{fig:err-single} we compare the result of using Patlak and TV-Patlak for estimating the uptake rates with respect to no noise, $20\%$ noise in the input function, Poisson noise in the sinogram, and finally with respect to the case in which the irreversibility  assumption is violated but without noise in the sinogram or input data. In each case the histogram of the relative errors is given on the left, the Patlak image in the middle and the TV-Patlak on the right. The different scales in the histograms are due to the total number of results illustrated. When there is no noise (triples $000$ and $+00$) the histogram illustrates results over all voxels but only one simulation, while for the noisy simulations the results are for all voxels over all $100$ realizations of the noise.  The TV-Patlak images are more homogeneous in all cases and the relative errors are smaller. The figures clearly show the improvements
of employing the TV-Patlak method as compared to using Patlak independently for each voxel.  This is confirmed in figure~\ref{fig:err-general} in which images with noise in the sinogram, positive $k_4$  and different noise levels in the input function are shown.

\begin{figure}[bp]
\centerline{
\includegraphics[height=3in,width=2.8in]{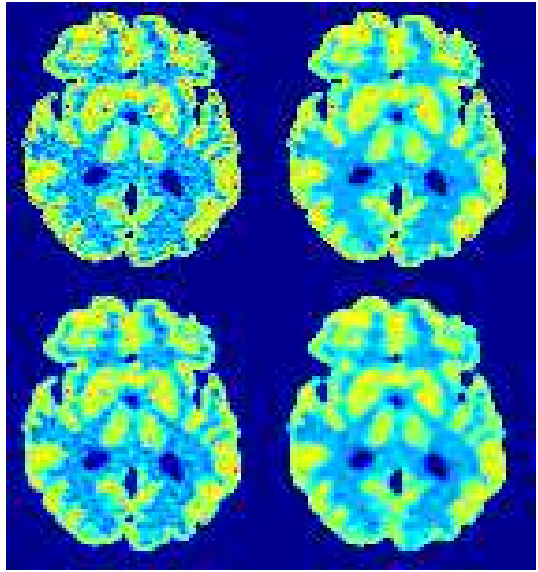}
}

\caption{Comparison of Patlak (upper-left), TV-Patlak (upper-right), Patlak-GF (bottom-left) and Patlak-MF (bottom-right) for imaging the uptake rate $K$ for one simulated data case $+4+$, i.e. $k_4>0$, $20\%$ noise in the input function and Poisson noise in the sinogram.\label{fig:img-4methods}}
\end{figure}

\begin{figure}[bp]
\centerline{
\includegraphics[scale=0.5]{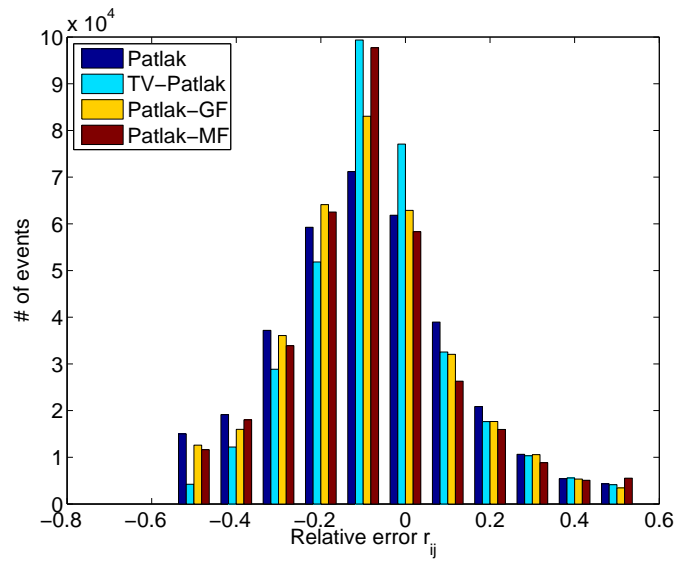}
}
\caption{Histograms for the relative error (\ref{eq:relerr}) in the uptake rate  calculated by Patlak, TV-Patlak, Patlak-GF and Patlak-MF, for
$100$ realizations of the data case $+4+$, i.e. $k_4>0$, $20\%$ noise in the input function and Poisson noise in the sinogram.\label{fig:hist-4methods}}
\end{figure}

Quantitative measurements, confirming the illustrations,  are presented in  table~\ref{tab:P2TVP-err-mod2}. There we also present the results for  conventional Patlak's method with  post-smoothing by two standard filters:
\begin{enumerate}
\item a Gaussian filter (Patlak-GF),  size $3$-by-$3$ with standard deviation $0.5$, generated by Matlab functions \texttt{filter2}(\texttt{fspecial}(`\texttt{gaussian}', $3$, $0.5$), \texttt{img}) and 
\item a median filter (Patlak-MF) generated by Matlab function \texttt{medfilt2(\texttt{img})} for a $3$-by-$3$ neighborhood.
\end{enumerate} 
Consistent with the observation in \cite{Phelps79,lammertsma1987}, we find that violation of the Patlak assumption, $k_4=0$, introduces about $10\%$ bias; $\bar{r}\approx 0$
when $k_4=0$ but $\bar{r}\approx 10\%$ for  $k_4>0$.  
The rows $\overline{TV}$ (std) and  ``\# 10\% (\#15\%)"
provide complementary supporting information, indicating that the TV is minimized by TV-Patlak; as  compared to  Patlak, Patlak-GF and Patlak-MF  the number of voxels with larger error is reduced. In particular, we emphasize that {\bf  TV-Patlak provides a better noise removal mechanism than  popular post-filtering approaches}. 

In figures~\ref{fig:img-4methods} and \ref{fig:hist-4methods} we illustrate the uptake rates and relative error in the uptake rates, respectively, calculated by Patlak, TV-Patlak, Patlak-GF and Patlak-MF for one simulated data case $+4+$, i.e. $k_4>0$, $20\%$ noise in the input function and Poisson noise in the sinograms. The uptake rate image generated by Patlak-MF is  visually smoother than that by TV-Patlak, but the equivalent histograms show that the relative error is  higher for Patlak-MF than for TV-Patlak; the Patlak-MF image is over-smoothed.

\begin{figure}[bp]
\centerline{
\includegraphics[scale=0.5]{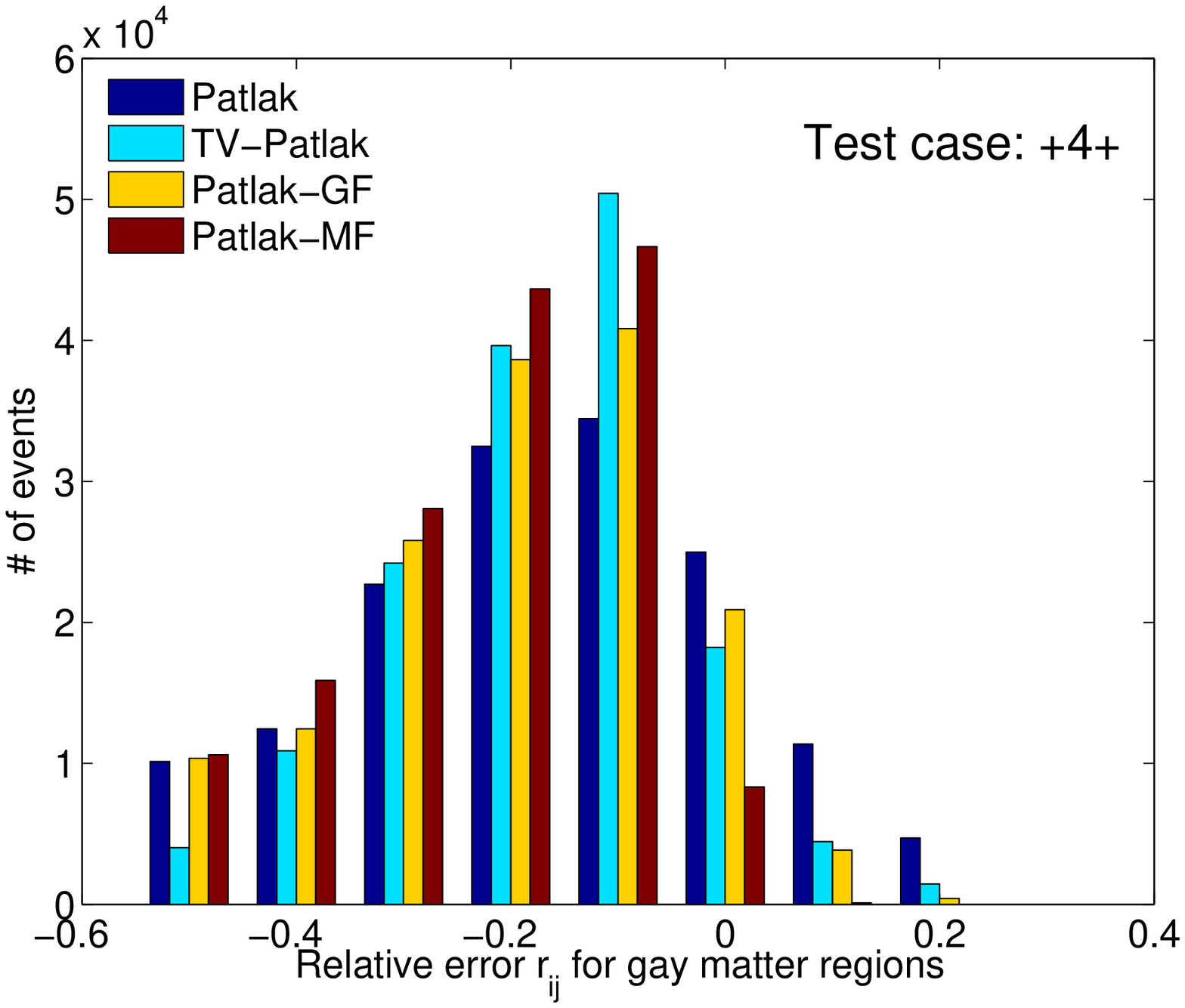}
}
\caption{Histograms for the relative error (\ref{eq:relerr}) in the uptake rate of gray matter regions calculated by Patlak, TV-Patlak, Patlak-GF and Patlak-MF, for
$100$ realizations of the data case $+4+$, i.e. $k_4>0$, $20\%$ noise in the input function and Poisson noise in the sinogram.\label{fig:hist-4methods-gray}}
\end{figure}

In figure~\ref{fig:hist-4methods-gray} we illustrate  the relative error of noisy case $+4+$ for gray matter regions  in the phantom, including frontal lobes, occipital lobes, insula cortex, temporal lobes and globus pallidus, which are of interest for Alzheimer's disease research. The error bars show that 
all estimation methods for gray matter regions have negative bias and  there are fewer cases with high error by TV-Patlak.

\begin{table}[htbp]
\begin{center}
\caption{
Results of phantom simulation: comparison of relative errors between the  Patlak (P), TV-Patlak (TV-P), Patlak-GF (P-GF) and Patlak-MF (P-MF) methods. $\overline{TV}$ is the mean of the total variation over $100$ simulations in each case, std denotes its standard deviation. The rows   `` \#10\%(\#15\%)" report the average number of voxels with high errors, $|r_{ik}|>10\%$
($|r_{ik}|>15\%$) per image. The first row indicates the specific simulation.
\label{tab:P2TVP-err-mod2}}

\scriptsize
\begin{tabular}{|c|c|c|c|c|c|c|c|c|}
\hline
\multicolumn{2}{|c|}{Data case}&  $000$ &  $040$ &  $00+$ &  $+00$ &  $+1+$ &  $+2+$ &  $+4+$ \\ \hline 
& $\bar{r}$ &  $-0.009$ &  $0.005$ &  $-0.009$ &  $-0.105$ &  $-0.105$ &  $-0.103$ &  $-0.088$ \\ \cline{3-9} 
& $d$ &  $0.116$ &  $0.136$ &  $0.157$ &  $0.106$ &  $0.148$ &  $0.151$ &  $0.165$ \\ \cline{2-9} 
& $\Bar{R}$ &  $0.116$ &  $0.136$ &  $0.157$ &  $0.147$ &  $0.176$ &  $0.177$ &  $0.182$ \\ \cline{3-9} 
P& $D$ &  $0.088$ &  $0.094$ &  $0.101$ &  $0.081$ &  $0.104$ &  $0.106$ &  $0.113$ \\ \cline{2-9} 
& $\overline{TV}$ &  $19.488$ &  $22.816$ &  $24.811$ &  $17.246$ &  $22.901$ &  $23.526$ &  $26.074$ \\ \cline{3-9} 
& std &  $0.000$ &  $0.193$ &  $0.235$ &  $0.000$ &  $0.219$ &  $0.225$ &  $0.301$ \\ \cline{2-9} 
& \#10\% &  $1429$ &  $1718$ &  $2020$ &  $2099$ &  $2272$ &  $2265$ &  $2238$ \\ \cline{3-9} 
& \#15\% &  $ 995$ &  $1166$ &  $1453$ &  $1285$ &  $1733$ &  $1732$ &  $1718$ \\ \hline 
& $\bar{r}$ &  $-0.005$ &  $0.020$ &  $0.001$ &  $-0.101$ &  $-0.094$ &  $-0.088$ &  $-0.063$ \\ \cline{3-9} 
& $d$ &  $0.104$ &  $0.120$ &  $0.128$ &  $0.097$ &  $0.121$ &  $0.123$ &  $0.133$ \\ \cline{2-9} 
T& $\Bar{R}$ &  $0.105$ &  $0.121$ &  $0.128$ &  $0.145$ &  $0.157$ &  $0.155$ &  $0.148$ \\ \cline{3-9} 
V& $D$ &  $0.095$ &  $0.096$ &  $0.099$ &  $0.078$ &  $0.089$ &  $0.091$ &  $0.096$ \\ \cline{2-9} 
$|$& $\overline{TV}$ &  $15.663$ &  $17.624$ &  $15.484$ &  $13.650$ &  $13.607$ &  $13.988$ &  $15.344$ \\ \cline{3-9} 
P& std &  $0.000$ &  $0.156$ &  $0.220$ &  $0.000$ &  $0.241$ &  $0.254$ &  $0.251$ \\ \cline{2-9} 
& \#10\% &  $1345$ &  $1462$ &  $1522$ &  $1905$ &  $2164$ &  $2097$ &  $1923$ \\ \cline{3-9} 
& \#15\% &  $ 974$ &  $1046$ &  $1102$ &  $1163$ &  $1441$ &  $1415$ &  $1348$ \\ \hline 
& $\bar{r}$ &  $-0.012$ &  $0.002$ &  $-0.012$ &  $-0.108$ &  $-0.108$ &  $-0.106$ &  $-0.091$ \\ \cline{3-9} 
& $d$ &  $0.124$ &  $0.137$ &  $0.151$ &  $0.115$ &  $0.142$ &  $0.143$ &  $0.151$ \\ \cline{2-9} 
P& $\Bar{R}$ &  $0.125$ &  $0.137$ &  $0.152$ &  $0.162$ &  $0.175$ &  $0.175$ &  $0.173$ \\ \cline{3-9} 
$|$& $D$ &  $0.106$ &  $0.106$ &  $0.104$ &  $0.088$ &  $0.101$ &  $0.102$ &  $0.106$ \\ \cline{2-9} 
G& $\overline{TV}$ &  $19.052$ &  $20.635$ &  $22.254$ &  $16.881$ &  $20.231$ &  $20.502$ &  $21.653$ \\ \cline{3-9} 
F& std &  $0.000$ &  $0.126$ &  $0.171$ &  $0.000$ &  $0.147$ &  $0.155$ &  $0.201$ \\ \cline{2-9} 
& \#10\% &  $1492$ &  $1589$ &  $1874$ &  $2240$ &  $2297$ &  $2279$ &  $2205$ \\ \cline{3-9} 
& \#15\% &  $1183$ &  $1191$ &  $1342$ &  $1328$ &  $1722$ &  $1713$ &  $1658$ \\ \hline 
& $\bar{r}$ &  $-0.017$ &  $-0.002$ &  $-0.021$ &  $-0.112$ &  $-0.116$ &  $-0.113$ &  $-0.095$ \\ \cline{3-9} 
& $d$ &  $0.125$ &  $0.138$ &  $0.148$ &  $0.117$ &  $0.139$ &  $0.140$ &  $0.147$ \\ \cline{2-9} 
P& $\Bar{R}$ &  $0.128$ &  $0.138$ &  $0.150$ &  $0.169$ &  $0.182$ &  $0.181$ &  $0.176$ \\ \cline{3-9} 
$|$& $D$ &  $0.113$ &  $0.115$ &  $0.114$ &  $0.096$ &  $0.103$ &  $0.104$ &  $0.109$ \\ \cline{2-9} 
M& $\overline{TV}$ &  $15.938$ &  $16.295$ &  $16.795$ &  $14.104$ &  $15.026$ &  $15.121$ &  $15.506$ \\ \cline{3-9} 
F& std &  $0.000$ &  $0.119$ &  $0.130$ &  $0.000$ &  $0.137$ &  $0.143$ &  $0.166$ \\ \cline{2-9} 
& \#10\% &  $1499$ &  $1522$ &  $1709$ &  $2342$ &  $2396$ &  $2362$ &  $2215$ \\ \cline{3-9} 
& \#15\% &  $1118$ &  $1152$ &  $1251$ &  $1335$ &  $1730$ &  $1710$ &  $1615$ \\ \hline

 \end{tabular}
\normalsize
\end{center}
\end{table}

%
%
%
%
%
%
%

Finally, we note that the computational cost for the TV-Patlak Algorithm \ref{alg:tikTV} is about $6.5$ seconds while for Patlak, Patlak-GF and Patlak-MF they are $0.48, 0.55$ and $0.58$ seconds respectively for a $2$D image on a PC with 1GHz CPU and Matlab code.

\section{Discussion} \label{sec:dis}
In this section we discuss issues relevant to the proposed TV-Patlak method. 
\begin{enumerate}
\item {\it Regularization constant $\alpha$:} There are many approaches for determining the choice of an appropriate regularization constant $\alpha$. A good reference would be \cite{Vogelbook02} in which the methods of unbiased predictive risk-estimation, generalized cross validation,  and  the L-curve are described. In general, the choice of $\alpha$ depends on the noise level of the problem. 
Here $\alpha$  balances the homogeneity of the uptake rate  against the residual of the traditional Patlak least squares data fit.  For the PET imaging application, noise in the TTAC data depends on the scanner, the  reconstruction method, the tracer dosage and even the kinetics of individuals.
It is therefore possible to make a standard parameter setting for commonly-used environments.  The most convenient method for the selection of $\alpha$ is the so-called L-curve, \cite{Hansen92,Hansen93}, which plots total variation against $\sum_{i=1}^N \|W_i(A (x_i,y_i)^T-\bfb_i) \|_2^2$ for all tested $\alpha$. The L-curve clearly displays the compromise between of the homogeneity and the residual of the  Patlak fitting equations. The $\alpha$ corresponding to the left lower corner of the L-curve is considered as the optimal choice.  One representative L-curve of our simulations is illustrated in figure~\ref{fig:Lcurve}. We found for our simulations that a suitable choice is $\alpha=0.2$, but certainly it will in general depend on the reconstruction algorithm. For example, the simple EM method and filtered backprojection algorithms introduce  more noise than the  maximum a posteriori (MAP) algorithm, \cite{Geman85,Bouman96}. For real data not only are there  additional sources of noise but the choice for $\alpha$ will also depend on the specific tracer. However, once an appropriate $\alpha$ is found by L-curve for a specific imaging environment, it can be fixed for future imaging calculations.

\begin{figure}[bp]
\centerline{\includegraphics[height=4in,width=4in]{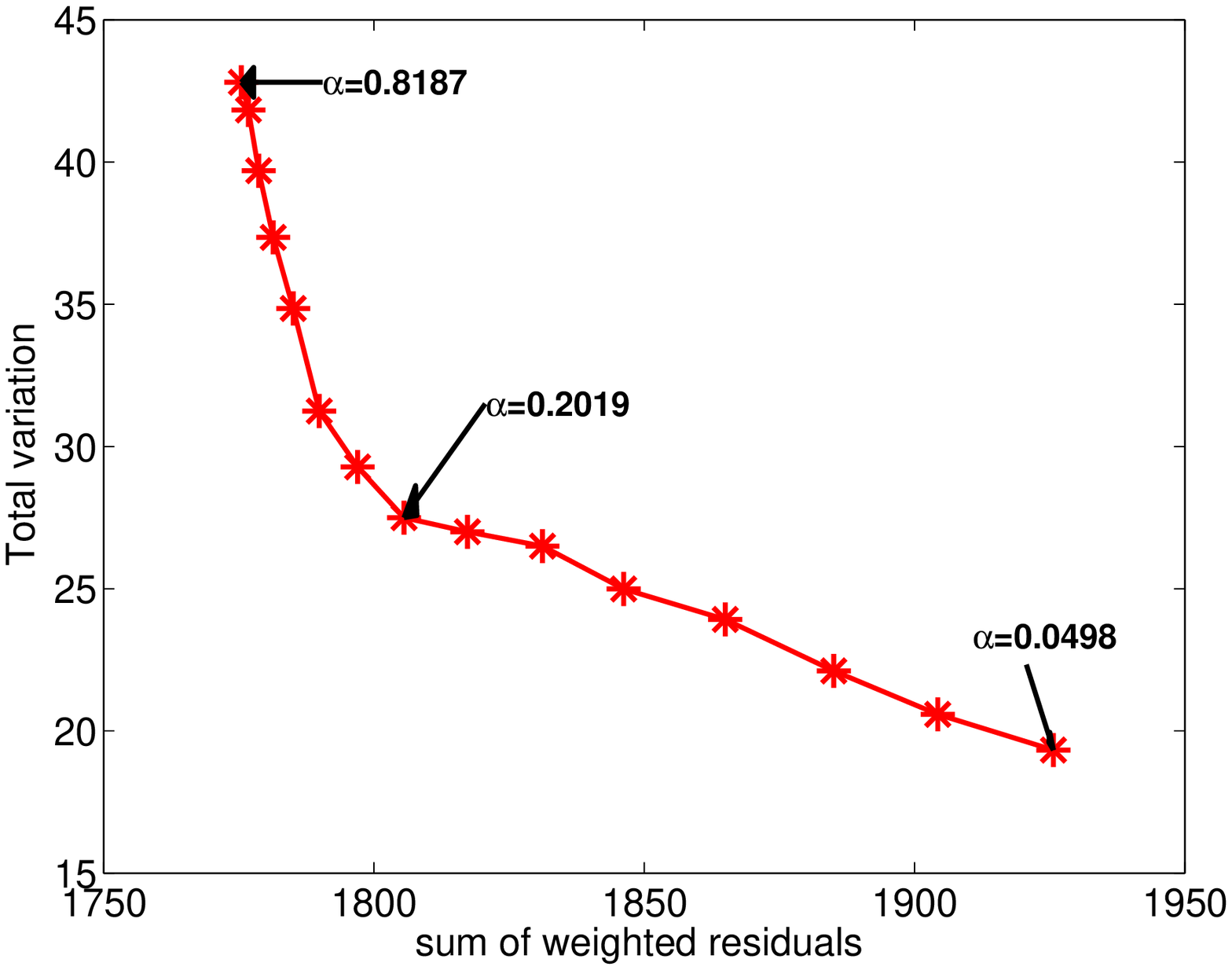}}
 \caption{L-curve for the simulation case `+4+'. Total variation against the sum of weighted residuals, i.e. $\sum_{i=1}^N \|W_i(A (x_i,y_i)^T-\bfb_i) \|_2^2$, for $\alpha$ from $0.0498$ to $0.8187$ is plotted. $\alpha=0.2$ is associated with the left lower ``corner''.\label{fig:Lcurve}}
\end{figure}

\item {Constant $\beta$:} We use Newton's method  for the case $\beta\neq 0$  to solve the optimization problem (\ref{tv-patlak}). Other methods for solving the TV problem with $\beta\neq 0$ are discussed in \cite{Vogelbook02}, which includes the primal-dual method \cite{Chetal}. A good choice of $\beta$  avoids numerical difficulties for small derivatives and provides a good approximation of the TV. For our tests we found that the results are not sensitive to the choice of $\beta$ and thus  suggest the use of  $\beta=10^{-8}$ as a good choice for FDG-PET brain imaging. If we wish to avoid the choice for $\beta$ 
the TV term in (\ref{tv-patlak}) can be reformulated as a set of linear constraints, and other algorithms are possible,  \cite{Boyd04optbook}.
\item {\it Boundary values:} At each iteration of the algorithm, after updating  $\bfx$ and before doing any function evaluation or other calculation, the \textit{boundary} voxels need to be updated. The value of each boundary voxel is set to the average of its active neighbors in four directions. 
\item {\it Computational efficiency:}
For computational efficiency and to minimize memory usage,  it is important to not only use sparse storage strategies for the relevant matrices but also to use appropriate algorithms for the solution of  the large scale sparse linear systems given by (\ref{stepupdate}). Here we use the QMR iteration, \cite{Freund91}. Moreover, if a general Newton's or quasi-Newton method with approximated Hessian were used, the cost in time and memory would be much more expensive because the approximated Hessian matrix is generally dense. In addition to achieving sparsity, the Hessian matrix is calculated accurately;  and the convergence should be faster. For our simulations convergence is achieved in  eleven iteration steps on average. The Matlab code of the TV-Patlak method can be downloaded from \url{http://math.asu.edu/~hongbin}.
\end{enumerate}

\section{Conclusions} \label{sec:conc}
A qualitative improvement in imaging of  PET uptake  can be achieved by using a global model, with  the total variation as a penalty term, to obtain the  voxel uptake rates. The resultant uptake images have spatial homogeneity over brain regions with similar kinetics and distinct edges  between brain regions that have different kinetics. It is statistically validated that the TV-Patlak significantly reduces the relative errors of the calculated parameters as compared with those generated by  Patlak's graphical method, and post-smoothing by  Gaussian and median filters.

\noindent\textbf{Acknowledgement}
This work was supported by grants from the state of Arizona, the NIH, R01 MH057899 and P30 AG19610 and the NSF DMS 0652833 and DMS 0513214. The authors thank Chi-Chuan Chen for providing the transition matrix, generated by the Monte Carlo method, which is used in the PET image reconstructions and Drs Bouman and Musfata for offering their MAP reconstruction code, which provides comparisons between reconstruction methods. We also acknowledge Dr. Laszlo Balkay and Dr. Jeffrey Fessler for providing PET simulator and PET reconstruction software, respectively.

\clearpage
\appendix

\section{Appendix}
To simplify notation, we assume that the  weighting  matrix in (\ref{tv-patlak}) has been absorbed into $A$ and $\bfb_i$.
To simplify the expressions in the objective function (\ref{eq:TVfunct})  we introduce  the vector function $\bff(\bfz)=(f_1(\bfz), f_2(\bfz),\cdots,f_{2N}(\bfz))^T$, where $f_i(\bfz)=\phi_k(\bfx)^2$ for $i=1\dots N$ and $f_i(\bfz)=\|A[x_{i-N};y_{i-N}]-\bfb_{i-N}\|^2$ for $i=N+1\dots 2N$.

\noindent\textbf{Gradient calculation} \\ 
To find the gradient of $\Phi(\bfz)$, we first derive the Jacobian of $\bff(\bfz)$. 

\begin{enumerate}
\item For $i=1,\cdots,N$  and $l=1,\cdots,2N$

\begin{eqnarray}\label{eq:Pf1}
\frac{\partial f_i}{\partial z_l}&=&\left\{\begin{array}{ll}
4x_i-2x_{i_b}-2x_{i_r}, & {\rm if\ } l=i \\
2x_{i_b}-2x_i,& {\rm if\ } l=i_b \\
2x_{i_r}-2x_i, & {\rm if\ } l=i_r \\
0, & {\rm otherwise. }
\end{array}\right. 
\end{eqnarray}

\item  For $i=N+1,\cdots,2N$  and $l=1,\cdots,2N$
  
\begin{equation}\label{eq:Pf2}
\frac{\partial f_i}{\partial z_l}=\left\{\begin{array}{ll}
p_{i1}, & {\rm if\ }l=i-N \\
p_{i2}, & {\rm if\ }l=i \\
     0, & {\rm otherwise},\end{array}\right .
\end{equation}
where 
$$\left (\begin{array}{c} p_{i1}\\p_{i2} \end{array}\right)=2A^TA \left (\begin{array}{c} x_{i-N} \\ y_{i-N} \end{array}\right)-2A^T\bfb_{i-N}.$$

Because $\nabla_{\bfx}\phi_i=1/(2\phi_i)\nabla_{\bfx} f_i$ for $i=1,\cdots,N$,  $\nabla_{\bfz}\Phi=(\nabla{\bff}(\bfz))^T\bfg,$ where
$$\bfg=[1/(2\phi_1), 1/(2\phi_2),\cdots,1/(2\phi_N),\alpha,\alpha,\cdots,\alpha]^T.$$
\end{enumerate}

\noindent{\textbf{Hessian matrix}}\\ 
To find the Hessian matrix for $\Phi(\bfz)$, we first derive the Hessian matrix for each function $f_i(\bfz)$. 

\begin{enumerate}
\item For $i=1,\cdots,N$, 
 $\nabla^2_{\bfx\bfx} f_i$ is  sparse
\begin{equation}\label{eq:Hf1}
\nabla^2_{\bfx\bfx} f_i=
\begin{array}{cl}
\begin{array}{rrr} i\ \ \ &i_b&\  i_r \end{array}& \\
\left (\begin{array}{rrr} 4&-2&-2\\-2&2&0\\-2&0&2\end{array}\right )&
\begin{array}{c}i\\i_b\\i_r\end{array}
\end{array}.
\end{equation}
\item For $i=N+1,\cdots, 2N$, the  Hessian matrix is again sparse. 
\begin{equation}\label{eq:Hf2}
\nabla^2_{\bfz\bfz} f_i=
	\begin{array}{cl}
\begin{array}{lr}{i-N} &i\end{array}& \\
\left (\begin{array}{cc}
q_{11}\quad \quad &q_{12}\\q_{21}\quad \quad &q_{22}
\end{array}\right )&
\begin{array}{c}i-N\\i,
\end{array}
\end{array}
\quad \mathrm{where} \quad
\left (\begin{array}{cc} q_{11}&q_{12}\\ q_{21} &q_{22} \end{array}\right)=2A^TA .\end{equation} 

Thus
$$\nabla^2\Phi(\bfz)=\sum_{i=1}^N \left ( \begin{array}{cc}
                    \nabla^2_{\bfx\bfx}\phi_i ,& 0_{N\times N} \\
                 0_{N\times N}, & 0_{N\times N}
               \end{array}\right )
  +\alpha \sum_{i=N+1}^{2N}\nabla^2 f_i,$$
where 
$$\nabla^2_{\bfx\bfx}\phi_i=-\frac{1}{4\phi_i^3}\nabla_{\bfx} f_i(\nabla_{\bfx} f_i)^T+ \frac{1}{2\phi_i}\nabla^2_{\bfx\bfx} f_i,$$

\end{enumerate}

\vspace{1cm}
{\bf Hongbin Guo} received his Ph.D. degree in computational mathematics from the Department of Mathematics, Fudan University, in 2000. He is currently an Assistant Research Professor with the Department of Mathematics and Statistics at Arizona State University. His main research interest is computational mathematics with a focus on the numerical linear algebra and its applications in medical imaging, particularly for the  quantification of brain function with  positron emission tomography and magnetic resonance imaging, co-registration, functional clustering, image restoration and classification.  \\

{\bf Rosemary A. Renaut}  received the  Ph.D. degree in applied mathematics from the University of Cambridge, 
U.K., in 1985. Since 1987, she has been with the Department of Mathematics, Arizona State University, where she is now a Full Professor. Dr. Renaut is a Fellow of the Institute for Mathematics and its Applications, and a Chartered Mathematician. Her research interests are broad and include the design and evaluation of computational methods for the solution of partial differential equations, with specific emphasis on high-order and spectral methods, as well as the development of novel algorithms for solving inverse problems, specifically as applied to medical image reconstruction and restoration.   \\

{\bf Kewei Chen} got his master degree from Beijing Normal University
in 1986 and his Ph.D. Degree from UCLA in 1993.  He is currently the
director and a senior biomathematician of the Computational Image
Analysis Program, Banner Alzheimer's Institute.  He has been using and
developing neuroimaging analytic techniques in PET, magnetic resonance
imaging (MRI) and functional MRI. His main methodological research
interests include the tracer kinetic modeling in PET, measuring local
and global volume changes using various MRI techniques, brain
functional connectivity, and multi-modal data integration. 
One of his primary research interests is the use of neuroimaging
techniques in the study of Alzheimer's disease.\\

{\bf Eric M Reiman} is Executive Director of the Banner Alzheimer's Institute,
Clinical Director of the Neurogenomics Division at the Translational
Genomics Research Institute (TGen), Professor and Associate Head of
Psychiatry at the University of Arizona, and Director of the Arizona
Alzheimer's Consortium. He received his undergraduate and medical
degrees and most of psychiatry residency training at Duke University. He
completed his residency training, became an Assistant Professor of
Psychiatry and developed a leadership role in positron emission
tomography research at Washington University in St. Louis, before moving
to Arizona. His research interests include brain imaging, genomics, and
their use in the unusually early detection and tracking of Alzheimer's
disease and the rigorous and rapid evaluation of promising Alzheimer's
disease-slowing, risk-reducing and prevention therapies.


\begin{thebibliography}{26}
\expandafter\ifx\csname natexlab\endcsname\relax\def\natexlab#1{#1}\fi
\expandafter\ifx\csname url\endcsname\relax
  \def\url#1{\texttt{#1}}\fi
\expandafter\ifx\csname urlprefix\endcsname\relax\def\urlprefix{URL }\fi

\bibitem[{Bouman and Sauer(1996)}]{Bouman96}
Bouman, C.~A., Sauer, K., March 1996. A unified approach to statistical
  tomography using coordinate descent optimization. IEEE Tr. Im. Proc. 5~(3),
  480--492.

\bibitem[{Boyd and Vandenberghe(2004)}]{Boyd04optbook}
Boyd, S., Vandenberghe, L., 2004. Convex Optimization. {Cambridge University
  Press}.

\bibitem[{Carson et~al.(1986)Carson, Huang, and Green}]{Carson:98}
Carson, R., Huang, S., Green, M., 1986. Weighted integration method for local
  cerebral blood flow measurements with positron emission tomography. J. Cereb.
  Blood Flow Metab. 6~(2), 245--58.

\bibitem[{Chan et~al.(1999)Chan, Golub, and Mulet}]{Chetal}
Chan, T.~F., Golub, G.~H., Mulet, P., 1999. A nonlinear primal-dual method for
  total variation-based image restoration. SIAM J. Sci. Comput. 20~(6),
  1964--1977.

\bibitem[{Chen et~al.(1998)Chen, Lawson, Reiman, Cooper, Feng, Huang, Bandy,
  Ho, Yen, and Palant}]{ChenGLLS:98}
Chen, K., Lawson, M., Reiman, E., Cooper, A., Feng, D., Huang, S., Bandy, D.,
  Ho, D., Yen, L., Palant, A., 1998. Generalized linear least squares method
  for fast generation of myocardial blood flow parametric images with
  \mbox{N}$-13$ ammonia \mbox{PET}. Med. Imag. 7~(3), 236--243.

\bibitem[{Feng and Huang(1996)}]{Feng96}
Feng, D., Huang, S., 1996. An unbiased parametric imaging algorithm for
  nonuniformly sampled biomedical system parameter estimation. IEEE Trans. Med.
  Imag. 15~(4), 512--518.

\bibitem[{Freund and Nachtigal(1991)}]{Freund91}
Freund, R.~W., Nachtigal, N.~M., 1991. {QMR}: {A} quasi-minimal residual method
  for non-{Hermitian} linear systems. Numerische Mathematik 60, 315--340.

\bibitem[{Geman and McClure(1985)}]{Geman85}
Geman, S., McClure, D.~E., 1985. Bayesian image analysis: An application to
  single photon emission tomography. In: Proc. Amer. Statist. Assoc.
  Statistical Computing Section. pp. 12--18.

\bibitem[{Guo et~al.(2007)Guo, Renaut, and Chen}]{Guoinpf07}
Guo, H., Renaut, R., Chen, K., 2007. An input function estimation method for
  \mbox{FDG-PET} human brain studies. Nuclear Medicine and Biology 34~(5),
  483--492.

\bibitem[{Hansen(1992)}]{Hansen92}
Hansen, P.~C., 1992. Analysis of discrete ill-posed problems by means of the
  \mbox{L}-curve. SIAM Review 34, 561--580.

\bibitem[{Hansen and O'Leary(1993)}]{Hansen93}
Hansen, P.~C., O'Leary, D.~P., 1993. The use of the \mbox{L}-curve in the
  regularization of discrete ill-posed problems. SIAM Journal on Scientific
  Computing 14, 1487--1503.

\bibitem[{Huang et~al.(1980)Huang, Phelps, Hoffman, Sideris, Selin, and
  Kuhl}]{Huetal:80}
Huang, S.-C., Phelps, M.~E., Hoffman, E.~J., Sideris, K., Selin, C.~J., Kuhl,
  D.~E., 1980. Noninvasive determination of local cerebral metabolic rate of
  glucose in man. Am. J. Physiol. 238~(E), 69--82.

\bibitem[{Jonsson et~al.(1998)Jonsson, Huang, and Chan}]{Jonsson98}
Jonsson, E., Huang, S.~C., Chan, T., 1998. Total variation regularization in
  positron emission tomography. Technical Reports on Image Processing CAM
  98-48, UCLA.

\bibitem[{Kaufman(1987)}]{Kaufman1987}
Kaufman, L., Mar. 1987. Implementing and accelerating the \mbox{EM} algorithm
  for positron emission tomography. IEEE Trans. Med. Imag. 6~(1), 37--51.

\bibitem[{Kisilev et~al.(2001)Kisilev, Zibulevsky, and Zeevi}]{KisilevZZ01}
Kisilev, P., Zibulevsky, M., Zeevi, Y.~Y., 2001. Wavelet representation and
  total variation regularization in emission tomography. In: ICIP (1). pp.
  702--705.

\bibitem[{Lammertsma et~al.(1987)Lammertsma, Brooks, Frackowiak, Beaney,
  Herold, Heather, Palmer, and Jones}]{lammertsma1987}
Lammertsma, A., Brooks, D., Frackowiak, R., Beaney, R., Herold, S., Heather,
  J., Palmer, A., Jones, T., 1987. Measurement of glucose utilisation with
  [18f]2-fluoro-2-deoxy-d-glucose: a comparison of different analytical
  methods. J. Cereb. Blood Flow Metab. 7~(2), 161--72.

\bibitem[{Logan et~al.(2001)Logan, Fowler, Volkow, Ding, Wang, and
  Alexoff}]{logan2001a}
Logan, J., Fowler, J., Volkow, N., Ding, Y., Wang, G., Alexoff, D., 2001. A
  strategy for removing the bias in the graphical analysis method. J. Cereb.
  Blood Flow Metab. 21~(3), 307--20.

\bibitem[{Logan et~al.(1990)Logan, Fowler, Volkow, Wolf, Dewey, Schlyer,
  MacGregor, Hitzemann, Bendriem, and Gatley}]{Logan90}
Logan, J., Fowler, J.~S., Volkow, N.~D., Wolf, A.~P., Dewey, S.~L., Schlyer,
  D.~J., MacGregor, R.~R., Hitzemann, R., Bendriem, B., Gatley, S.~J., 1990.
  Graphical analysis of reversible radioligand binding from time-activity
  measurements applied to [\mbox{N}-11\mbox{C}-methyl]-(-)-cocaine \mbox{PET}
  studies in human subjects. J. Cereb. Blood Flow Metab. 10, 740--747.

\bibitem[{Matlab(2008)}]{matlab}
Matlab, 2008. Matlab is a registered trademark of \mbox{MathWorks, Inc.}
\newline\urlprefix\url{http://www.mathworks.com}

\bibitem[{Patlak et~al.(1983)Patlak, Blasberg, and Fenstermacher}]{Patlak83}
Patlak, C.~S., Blasberg, R.~G., Fenstermacher, J.~D., 1983. Graphical
  evaluation of blood-to-brain transfer constants from multiple-time uptake
  data. J. Cereb. Blood Flow Metab. 3~(1), 1--7.

\bibitem[{Phelps et~al.(1979)Phelps, Huang, Hoffman, Selin, and
  Kuhl}]{Phelps79}
Phelps, M.~E., Huang, S.-C., Hoffman, E.~J., Selin, C.~E., Kuhl, D., 1979.
  Tomographic measurement of local cerebral glucose metabolic rate in man with
  (\mbox{$^{18}$F}) fluorodeoxyglucose: \mbox{V}alidation of method. Ann.
  Neurol. 6, 371--388.

\bibitem[{Rudin et~al.(1992)Rudin, Osher, and Fatemi}]{Ruetal}
Rudin, L., Osher, S., Fatemi, E., 1992. Nonlinear total variation based noise
  removal algorithms. Physica D 60, 259--268.

\bibitem[{Sokoloff et~al.(1977)Sokoloff, Reivich, Kennedy, Rosiers, Patlak,
  Pettigrew, Sakurada, and Shinohara}]{Sokoloffs}
Sokoloff, L., Reivich, M., Kennedy, C., Rosiers, M. H.~D., Patlak, C.~S.,
  Pettigrew, K.~D., Sakurada, M., Shinohara, M., 1977. The [$^{14}$\mbox{C}]
  deoxyglucose method for the measurement of local cerebral glucose metabolism:
  theory procedures and normal values in the conscious and anesthetized albino
  rat. J. Neurochem. 28, 897--916.

\bibitem[{Vogel(2002)}]{Vogelbook02}
Vogel, C.~R., 2002. Computational Methods for Inverse Problems. Society for
  Industrial and Applied Mathematics, Philadelphia, PA, USA.

\bibitem[{Zhou et~al.(2003)Zhou, Endres, Brasic, Huang, and Wong}]{Zhou2003}
Zhou, Y., Endres, C.~J., Brasic, J.~R., Huang, S.-C., Wong, D.~F., 2003. Linear
  regression with spatial constraint to generate parametric images of
  ligand-receptor dynamic \mbox{PET} studies with a simplified reference tissue
  model. NeuroImage 18~(4), 975--989.

\bibitem[{{Zubal} et~al.(1994){Zubal}, {Harrell}, {Smith}, {Rattner}, {Gindi},
  and {Hoffer}}]{Zubalphantom}
{Zubal}, I.~G., {Harrell}, C.~R., {Smith}, E.~O., {Rattner}, Z., {Gindi}, G.,
  {Hoffer}, P.~B., Feb. 1994. {Computerized three-dimensional segmented human
  anatomy}. Medical Physics 21, 299--302.

\end{thebibliography}
\end{document}